\documentclass[conference]{IEEEtran}

\usepackage{cite}
\usepackage{amsmath,amssymb}
\usepackage{graphicx}
\usepackage{xspace}
\usepackage{comment}
\usepackage{epigraph}
\usepackage{amsfonts}
\usepackage{soul}
\usepackage{mathtools, nccmath}
\usepackage{mathrsfs}
\usepackage{algorithm}
\usepackage{algpseudocode}
\usepackage[utf8]{inputenc}
\usepackage[T1]{fontenc}
\usepackage{siunitx}
\usepackage{multirow}
\usepackage{enumitem}
\usepackage{tikz}
\usepackage{verbatim}
\usetikzlibrary{calc}
\usepackage{soul}
\usepackage{adjustbox}

\usepackage{comment}

\algnewcommand\algorithmicswitch{\textbf{switch}}
\algnewcommand\algorithmiccase{\textbf{case}}
\algnewcommand\algorithmicassert{\texttt{assert}}
\algnewcommand\Assert[1]{\State \algorithmicassert(#1)}%
\algdef{SE}[SWITCH]{Switch}{EndSwitch}[1]{\algorithmicswitch\ #1\ \algorithmicdo}{\algorithmicend\ \algorithmicswitch}%
\algdef{SE}[CASE]{Case}{EndCase}[1]{\algorithmiccase\ #1}{\algorithmicend\ \algorithmiccase}%
\algtext*{EndSwitch}%
\algtext*{EndCase}%
\usepackage{hyperref}
\hypersetup{hidelinks}
\usepackage{amsthm}

\newtheorem{theorem}{Theorem}%
\newtheorem{proposition}{Proposition}%

\usepackage{ragged2e}
\usepackage{wrapfig}
\usepackage{stmaryrd}

\usepackage[most]{tcolorbox}
\newtcolorbox{myboxi}[1][]{
  breakable,
  title=#1,
  colback=gray!10!white,
  colbacktitle=white,
  coltitle=black,
  fonttitle=\bfseries,
  bottomrule=0.5pt,
  toprule=0.5pt,
  leftrule=0.5pt,
  rightrule=0.5pt,
  titlerule=0pt,
  colframe=black,
  boxsep=1pt,left=2pt,right=2pt,top=2pt,bottom=2pt
}

\usepackage{listings}
\usepackage{makecell}

\usepackage{multirow}
\usepackage{booktabs}
\usepackage{xcolor}
\usepackage{array}

\usepackage{wrapfig}

\definecolor{codegreen}{rgb}{0,0.6,0}
\definecolor{codegray}{rgb}{0.5,0.5,0.5}
\definecolor{codepurple}{rgb}{0.58,0,0.82}
\definecolor{backcolour}{rgb}{0.95,0.95,0.92}

\lstdefinestyle{mystyle}{
  backgroundcolor=\color{backcolour},
  commentstyle=\color{codegreen},
  keywordstyle=\color{magenta},
  numberstyle=\tiny\color{codegray},
  stringstyle=\color{codepurple},
  basicstyle=\ttfamily\footnotesize,
  breakatwhitespace=false,
  breaklines=true,
  captionpos=b,
  keepspaces=true,
  numbersep=5pt,
  showspaces=false,
  showstringspaces=false,
  showtabs=false,
  tabsize=2
}

\newcommand{\hippasus}[0]{\textsc{Hippasus}\xspace}
\def\ojoin{\setbox0=\hbox{$\bowtie$}\rule[-.02ex]{.25em}{.4pt}\llap{\rule[\ht0]{.25em}{.4pt}}}
\newcommand{\lojoin}{\mathbin{\ojoin\mkern-5.8mu\bowtie}}

\usepackage{tikz}

\begin{document}

\title{\hippasus: Accurate and Cost-Efficient Feature Augmentation over Relational Tables}

\author{\IEEEauthorblockN{Serafeim Papadias}
\IEEEauthorblockA{\textit{Athena Research Center}\\
Athens, Greece\\
serafeim.papadias@athenarc.gr}
\and
\IEEEauthorblockN{Kostas Patroumpas}
\IEEEauthorblockA{\textit{Athena Research Center}\\
Athens, Greece\\
kpatro@athenarc.gr}
\and
\IEEEauthorblockN{Dimitrios Skoutas}
\IEEEauthorblockA{\textit{Athena Research Center}\\
Athens, Greece\\
dskoutas@athenarc.gr}
}

\maketitle

\begin{abstract}
ML models critically depend on feature quality, yet in real-world settings, useful features are often distributed across multiple relational tables rather than a single dataset. Feature augmentation addresses this problem by automatically discovering and joining additional tables to enrich a base table with predictive features. However, scaling feature augmentation to complex schemas with many tables and multi-hop relationships is challenging. It requires exploring a large space of join paths, executing costly joins, and selecting useful features from noisy results. Existing approaches suffer from either limited effectiveness or efficiency. Restricting exploration to simple joins limits predictive performance, while more expressive methods rely on expensive training data, lack scalability, or fail to fully exploit schema-level semantics. We present \hippasus, a cost-aware, LLM-guided feature augmentation framework over relational schemas, that addresses these challenges. \hippasus ranks candidate join paths before executing any join, by combining lightweight statistical signals with adaptive semantic analysis, invoking an LLM only when necessary. It further introduces efficient multi-way join execution with cross-path feature consolidation, and a hybrid feature selection strategy that integrates statistical relevance with semantic refinement. 
Experiments on 13 real-world datasets show that \hippasus improves predictive performance by up to 15.7\% on average over
state-of-the-art methods, while running on par with the fastest of them and keeping its LLM token cost low.
\end{abstract}


\sloppypar

\section{Introduction}
\label{sec:introduction} 

\noindent Machine learning (ML) models are only as good as their input features, and the features that matter are often scattered across many relational tables rather than stored with the labels. For instance, the quantity of an order line may depend on the category and the supplier of the product, which are stored in other tables, reachable only through the product table. \emph{Feature augmentation}~\cite{vldb20-arda, icde22-autofeature, icde23-metam, icde24-autofeat, icde25-featpilot} automates this enrichment: given a \emph{base} table with labeled instances for a classification or regression task and a set of \emph{candidate} tables, it finds the \emph{join paths} that lead to useful attributes and adds these attributes to the base table as features, via possibly multi-hop joins.

Doing this both effectively and efficiently is hard. The number of candidate join paths grows exponentially with their length; judging a path by executing its joins, and often by training a model on the result, is expensive; and the joins bring in many noisy or redundant features, some of them reachable through several paths of different quality.

Existing approaches address these challenges only partially. ARDA~\cite{vldb20-arda} and METAM~\cite{icde23-metam} restrict exploration to one-hop joins, thus missing features that lie further away. AutoFeature~\cite{icde22-autofeature} and FeatPilot~\cite{icde25-featpilot} guide multi-hop exploration with reinforcement learning and LSTMs, respectively, both of which require task-specific training, while AutoFeat~\cite{icde24-autofeat} ranks features purely by statistical relevance and redundancy. 
FeatPilot also embeds columns with a pre-trained language model, but only to group similar ones. 
Yet all of them judge a join path by the data it produces, and none considers what the tables mean for the prediction task. 
The schema, however, often reveals which tables can matter: in the example above, the names \textit{categories} and \textit{suppliers} alone suggest that these tables may help predict order quantities.

We present \hippasus,\footnote{\url{https://github.com/spapadias/hippasus}} a feature augmentation framework built on this observation: whether a join path is worth executing can largely be judged before executing it. Language models (LMs) can read the schema, i.e., the names of tables and columns and their descriptions, together with the prediction task, and judge which tables are relevant, while statistics computed without joining estimate how well each hop connects. \hippasus therefore ranks the candidate paths without materializing any join, executes only the top-ranked ones, and selects features from their results. LMs already provide useful semantic signals for entity resolution~\cite{icde24-llm-re}, column type annotation~\cite{arxiv24-racoon, vldb24-archetype}, dataset description~\cite{autoddg-sigmod2026}, and table understanding~\cite{sigmod24-table-gpt, vldb24-chorus}, so using them to read a schema is natural. Making their judgments decide which joins run, at low cost, is not. To the best of our knowledge, \hippasus is the first feature augmentation approach to do so. It needs no fine-tuning and trains no model to explore join paths.

Since calls to large language models (LLMs) can be costly, \hippasus uses LMs adaptively. If an LLM cannot tell what the columns represent, \hippasus relies on statistics alone. Otherwise, it scores the tables with a small language model (SLM) and calls the LLM only if the SLM's scores do not clearly separate the most relevant tables from the rest.

Three further techniques complete the design. Cryptic column names obscure what the tables mean. 
To tackle this, an LLM first describes the columns, offline and independently of the task, and \hippasus relies on statistics alone only if the LLM still cannot tell what the columns represent. Joins must not drop labeled instances, so \hippasus executes the top-ranked paths with a variant of the Yannakakis algorithm~\cite{vldb81-yannakakis} that discards non-contributing tuples early while keeping every row of the base table. Finally, since the joins bring in noisy and redundant features, \hippasus consolidates features reached through different paths, ranks them by mutual information on the training data, and lets the LLM refine the top of the ranking. Our main contributions are:

\begin{itemize}[leftmargin=*]
\item \textbf{Rank before join.} A feature augmentation framework that ranks multi-hop join paths by LM-based semantic relevance and join statistics before materializing any join, and executes only the most promising ones.
\item \textbf{Adaptive semantic scoring.} A cost-aware cascade that relies on statistics alone when the columns stay uninterpretable, and otherwise calls an LLM only if an SLM does not clearly separate the most relevant tables from the rest.
\item \textbf{Efficient materialization and selection.} A suffix-based Yannakakis join strategy that preserves the rows of the base table, cross-path feature consolidation, and a feature selector that combines mutual information with LLM refinement.
\item \textbf{Extensive evaluation.} A comparison with three state-of-the-art methods on 13 real-world datasets, with component ablations and analyses of the cascade and the LLM choice.
\end{itemize}

In our experiments, \hippasus improves predictive performance by 15.7\%, 8.1\%, and 10.8\% on average over ARDA, AutoFeat, and FeatPilot, respectively, and takes 13.5 seconds per dataset on average, against 15.5 for AutoFeat, the fastest of them. Its cascade spends 46\% of the input tokens of always calling the LLM for an average accuracy loss of 1.1\%, and most of its gain on nine datasets comes from features two or more joins away.

The remainder of this paper proceeds as follows: Section~\ref{sec:related-work} discusses related work, Section~\ref{sec:problem-statement} defines the problem, Section~\ref{sec:components} presents \hippasus, Section~\ref{sec:experiments} evaluates it, and Section~\ref{sec:conclusions} concludes.

\section{Related Work}
\label{sec:related-work} 

\noindent \textbf{Feature Augmentation.}
Feature augmentation adds features from candidate tables to a base table to improve a downstream model; studies of when key--foreign key joins can be avoided without harming accuracy~\cite{sigmod16-kumar, vldb17-kumar} are complementary. Deep Feature Synthesis~\cite{dsaa15-kanter} explores join paths exhaustively and applies transformations along them, at a high computational cost. ARDA~\cite{vldb20-arda} improves efficiency by ranking candidate tables with discovery tools such as Aurum~\cite{icde18-aurum}, but remains limited to one-hop joins and model-agnostic heuristics. AutoFeature~\cite{icde22-autofeature} guides the exploration with multi-armed bandits and deep Q-networks, and METAM~\cite{icde23-metam} with multi-armed bandits in a goal-oriented feedback loop driven by downstream utility; both evaluate sampled join paths by executing joins and training models. AutoFeat~\cite{icde24-autofeat} avoids training by ranking multi-hop paths with relevance and redundancy metrics computed on the joined data, but ignores semantics. FeatPilot~\cite{icde25-featpilot} trains LSTM models to estimate path and feature utility and clusters columns by pre-trained language-model embeddings that ignore the prediction task. FeatAug~\cite{icde24-feataug} generates predicate-aware aggregation queries (e.g., averages) over a single one-to-many join between two tables, which is hard to extend to multi-hop paths. In contrast, \hippasus ranks multi-hop join paths before executing any join, using task-aware semantic signals obtained adaptively at controlled cost.

\noindent \textbf{Data and Feature Discovery with LLMs.}
LLMs increasingly support data discovery and annotation~\cite{DBLP:journals/debu/FreireFFKLPSSW25}: they annotate columns with semantic types without task-specific training~\cite{vldb24-archetype}, even without a predefined type set~\cite{arxiv26-stratyper}; describe datasets by combining data-driven summaries with LLM-based enrichment~\cite{autoddg-sigmod2026}; retrieve tables for natural-language questions by combining LLM-generated table representations with full-text and vector search~\cite{sigmod25-pneuma}; and synthesize the transformations that derive an augmented table from related tables~\cite{icde24-datalore}. None of them augments features for a prediction task; \hippasus's Feature Description Generator builds on this line of work to describe cryptic columns for its Path Explorer and Feature Selector. For single-table feature selection, LLMs rank features from their names and the task description, on par with LASSO and without training data~\cite{tmlrR25-llm-select}, set feature-specific LASSO penalties~\cite{DBLP:journals/corr/abs-2502-10648}, or are combined with random forests and sequential search~\cite{DBLP:journals/corr/abs-2503-24157} or refined by lightweight models such as decision trees~\cite{DBLP:journals/corr/abs-2412-08516}; text-based rankings that rely on feature descriptions have also been compared with data-driven ones that use data samples~\cite{DBLP:journals/sigkdd/LiTL24}. Decoupling feature selection from path exploration lets \hippasus benefit from such techniques: it extends LLM-Rank~\cite{DBLP:journals/sigkdd/LiTL24} with statistical signals computed on the augmented table. To the best of our knowledge, \hippasus is the first feature augmentation approach whose LM judgments decide which joins run.

\noindent \textbf{Joinable and Unionable Table Discovery.}
Joinable table discovery finds tables that join with a query table: Aurum~\cite{icde18-aurum} builds a knowledge graph of relationships between datasets, Lazo~\cite{icde19-lazo} and JOSIE~\cite{sigmod19-josie} scale overlap-based search to massive data lakes with locality-sensitive hashing and set similarity search, PEXESO~\cite{icde21-oyamada} and DeepJoin~\cite{vldb23-deepjoin} use embeddings to find semantic joins that tolerate misspellings and format differences, and Freyja~\cite{edbt25-freyja, tkde26-freyja} and OmniMatch~\cite{vldb25-omnimatch} learn join quality from data profiles and with graph neural networks, respectively; LakeBench~\cite{vldb24-lakebench} benchmarks join and union search on 16 million real tables. Correlated dataset search, such as COCOA~\cite{edbt21-cocoa} and QCR~\cite{icde22-qcr}, retrieves joinable tables whose columns correlate with a target column, using a rank-correlation index and correlation sketches, respectively; unlike \hippasus, these systems score single columns reached through one join and ignore what the columns mean. SemDisc~\cite{sigmod26-semdisc} combines equi-joins and semantic joins to recover multi-hop join paths whose results match user-provided example tuples. Unionable table search adds rows rather than features, using semantic relationships between column pairs~\cite{sigmod23-santos} or contextualized column embeddings~\cite{vldb23-starmie}. Joinable table discovery is complementary to \hippasus, as it can construct the join graph that \hippasus takes as input.

\section{Problem Definition}
\label{sec:problem-statement}

\begin{figure*}[t!]
\centering 
\includegraphics[width= \textwidth]{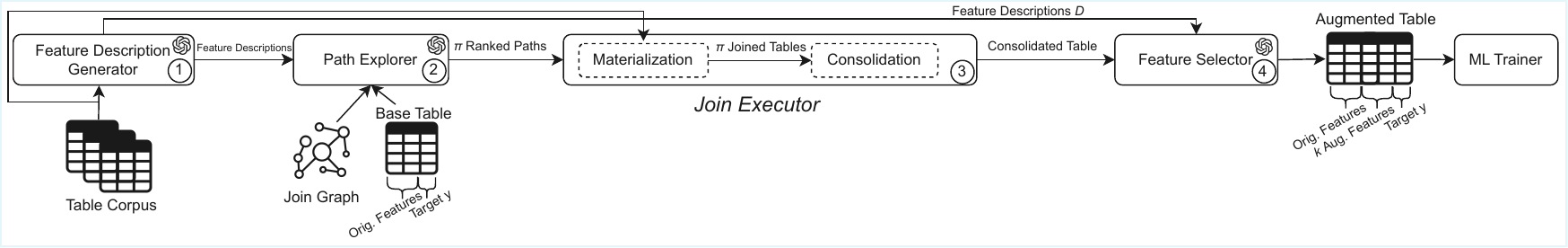}
\caption{Overview of \hippasus.} 
\label{fig:hippasus-overview} 
\end{figure*}

\noindent \textbf{ML Task.} 
A machine learning (ML) task seeks to learn a predictive function $f : \mathcal{X} \to \mathcal{Y}$ that maps an input feature space $\mathcal{X}$ to an output space $\mathcal{Y}$. Each input $\mathbf{x}_i \in \mathcal{X}$ represents a feature vector (corresponding to a record in a relational table), and each output $y_i \in \mathcal{Y}$ represents the target value for prediction. We consider two types of tasks: \textit{classification}, where $\mathcal{Y}$ is a discrete label space (binary or multiclass), and \textit{regression}, where $\mathcal{Y} \subseteq \mathbb{R}$ represents a continuous target domain. We focus on the standard setting of a single, predetermined target variable; supporting multiple prediction targets simultaneously is out of scope in this work.

\vspace{1mm} 
\noindent \textbf{Join Graph.} Consider a collection of \emph{relational tables} $\mathcal{D} = \{T_0, T_1, \ldots, T_n\}$ connected via a \emph{join graph} 
$G = (\mathcal{D}, E)$, where each node 
corresponds to a table $T_i \in \mathcal{D}$,
and each directed edge $e_{ij} \in E$ indicates that table $T_i$ contains a foreign key referencing the primary key of table $T_j$. Like previous works~\cite{vldb20-arda, icde22-autofeature, icde24-autofeat, icde25-featpilot}, we assume that primary-key/foreign-key (PK-FK) relationships are given as input, either via schema metadata or dataset discovery tools~\cite{icde18-aurum, vldb21-auctus}. 
Constructing the join graph via methods for joinability discovery~\cite{vldb23-deepjoin, sigmod19-josie, icde19-lazo, icde21-oyamada, tkde26-freyja, vldb25-omnimatch} is an orthogonal problem to \hippasus.
A \emph{join path} $p$ 
is a path in $G$ that represents an ordered sequence of join operations, each one referring to a valid relational join between two tables that share a joinable attribute (i.e., a PK-FK pair). 

\vspace{1mm}
\noindent \textbf{Base Table, Candidate Tables, and Augmented Table.} We denote by $T_{\text{base}}$ a table in $G$ 
that contains a set of attributes (i.e., features) and a target variable $y$ for prediction. The remaining tables in $G$ 
are referred to as \emph{candidate tables}. These tables can be potentially joined with $T_{\text{base}}$ via appropriate join paths consisting of consecutive edges defined in $G$ 
to augment it with additional features that may be helpful for predicting $y$. This augmented table, denoted as $T_{\text{aug}}$, is a table that includes all original attributes from $T_{\text{base}}$ along with additional attributes obtained by joining candidate tables via the join paths. The augmented table represents an enriched version of the base table whose feature space has been extended with attributes from the candidate tables. 
Without loss of generality, we assume that $T_0$ corresponds to the base table $T_{\text{base}}$ while $\{T_1, \ldots, T_n\}$ 
represent the candidate tables. 

\vspace{1mm} 
\noindent \textbf{Problem Statement.} 
Assume 
(a) a base table $T_{\text{base}}$ 
with a single target variable $y$,  
(b) the candidate tables $\mathcal{D} \setminus \{T_{\text{base}}\}$, connected to $T_{\text{base}}$ via a join graph $G$, and 
(c) a fixed downstream ML model $f$ together with a performance metric $\mathcal{M}$. 
The goal of \emph{feature augmentation} is to identify a set of join paths and a corresponding set of external features whose materialization yields an augmented table $T_{\text{aug}}$ that maximizes $\mathcal{M}(f, T_{\text{aug}})$, i.e., the predictive performance of $f$ on the augmented data. 

In practice, this objective is combinatorial and intractable in general, since it requires jointly searching over valid join paths and subsets of external features. 
Feature augmentation strictly generalizes classical feature subset selection, which is known to be NP-hard~\cite{welch1982algorithmic, DBLP:journals/ml/TsamardinosBKPC19, DBLP:conf/icml/JohnKP94, DBLP:journals/tnn/Battiti94}, by additionally introducing the search over candidate tables and join paths. 
Accordingly, as in prior work~\cite{vldb20-arda, icde24-autofeat, icde22-autofeature, icde25-featpilot, icde23-metam, dsaa15-kanter}, practical solutions rely on heuristic or approximate strategies to navigate this search space efficiently.

\section{Our Approach}
\label{sec:components}

As illustrated in Figure~\ref{fig:hippasus-overview}, \hippasus adopts a modular, pipeline-oriented architecture that progressively transforms the input schema into an augmented training table. Starting from the base and candidate tables, it first enriches schema metadata with semantic descriptions, then identifies promising join paths, materializes them efficiently, and finally selects the most informative features.

Concretely, the process begins with a \textbf{Feature Description Generator (FDG)}, which uses an LLM to produce compact, semantically meaningful descriptions of features and assess the availability of usable semantic signals. These descriptions are then leveraged by the \textbf{Path Explorer (PEX)}, which prioritizes promising join paths without materializing them, combining lightweight statistical indicators with adaptive semantic analysis to navigate the large search space efficiently. The selected paths are passed to the \textbf{Join Executor (JEX)}, which materializes them using multi-way joins and consolidates features originating from alternative paths into a unified representation. Finally, the \textbf{Feature Selector (FS)} produces the augmented table $T_{\text{aug}}$ by retaining features that are both statistically predictive and semantically relevant, while filtering out noisy or redundant ones.
The resulting table $T_{\text{aug}}$ is then used to \textbf{train downstream ML models} (e.g., Random Forests, Gradient Boosted Trees) via standard AutoML frameworks such as AutoGluon~\cite{arxiv20-autogluon-tabular}, providing an end-to-end evaluation of feature quality.

Next, we describe each component in detail.

\subsection{Feature Description Generator}
\label{sec:feature-description-generator} 

Real-world datasets often contain uninformative feature names 
that hinder reasoning over feature relevance and relationships across tables. 
FDG employs an LLM (prompt in Appendix~\ref{appendix:prompts} at the end of this paper) to automatically generate concise, semantically rich descriptions for each feature in the dataset's domain-specific context, enabling effective semantic analysis in subsequent steps.
FDG receives as input the names of the tables $\mathcal{D}$, the names of their features, and any available dataset descriptions (e.g., accompanying documentation in the form of a readme file). In the absence of such input, data samples and statistics can alternatively provide context to the LLM.
The aim is to enrich feature names that are not very informative (e.g., abbreviations) with descriptions that can enable or improve semantic analysis in subsequent steps in the pipeline.
Indicatively, in the datasets used in our experiments, \texttt{pupilDiamMax} is enriched with the description ``Maximum pupil diameter during fixation'', \texttt{V9} with ``Minimum luminosity value'', etc. 

In addition, in the same prompt, FDG asks the LLM to flag whether the feature names and descriptions carry sufficient \textit{semantic signal} for downstream semantic analysis.
This annotation constitutes the first layer of adaptivity in \hippasus: when both feature names and generated descriptions remain cryptic or semantically uninformative, downstream semantic scoring is disabled, and the system relies on statistical signals alone, avoiding noisy or misleading semantic judgments.
Since FDG does not depend on the target variable $y$ and the prediction task, it is executed offline as a preprocessing step.

\subsection{Path Explorer}
\label{sec:path-explorer} 

PEX takes as input the base table $T_{\text{base}}$, the join graph $G$, the target variable $y$, and a path budget $\pi$, and produces as output a ranked list of top-$\pi$ join paths $\mathcal{P} = \{p_1, p_2, \ldots, p_\pi\}$ most likely to yield high-quality features for the prediction task. 
We consider as candidate join paths all acyclic paths in $G$ starting from $T_{\text{base}}$ with length up to $\ell$, where the length of a path is its number of tables, so that $\ell=2$ covers one-hop paths ($\ell \in [2,7]$ in our experiments).

\hippasus conducts path exploration in two phases, which combine semantic understanding with statistical grounding: 
(1) \textit{semantic table scoring}, where semantic relevance is assessed adaptively for all candidate tables with respect to the target variable, and  
(2) \textit{hybrid path scoring}, which fuses the resulting semantic scores with statistical connection quality metrics during breadth-first search exploration. 
The rationale is that semantic analysis can identify domain-relevant features based on world knowledge, but lacks the data-specific signals needed to assess join feasibility. 

\subsubsection{Semantic Table Scoring} 
\label{subsec:semantic-table-scoring} 

\hippasus performs semantic table scoring adaptively.
It employs a \textit{cascade-style routing} policy inspired by prior work on model cascades and cost-aware semantic processing~\cite{vldb25-docetl, vldb25-lotus, sigmod26-bargain}.
It first consults FDG's semantic-signal annotation to determine whether usable semantic signal is available.
If not, semantic scoring is skipped, and all candidate tables receive a semantic score of zero. Otherwise, PEX scores candidates using an SLM and potentially an LLM, if the former is not deemed sufficient.

For SLM-based scoring, PEX constructs text representations for the base table and candidate tables from feature metadata.
Formally, let $\phi(T_i)$ denote the text representation of candidate table $T_i$, constructed from its feature metadata, namely feature names and FDG-generated descriptions, after excluding join-key columns. Let $\phi(T_{\text{base}}, y)$ denote the corresponding text representation of the base table and prediction target. 
Using a sentence embedding model, we compute a \emph{proxy semantic score} for each candidate table as follows: 
\begin{equation}
s_i^{\text{proxy}} = \cos\!\big(\text{Embed}(\phi(T_{\text{base}}, y)), \text{Embed}(\phi(T_i))\big),
\end{equation}
which is then \textit{normalized} to:
\begin{equation}
s_i^{\text{norm}} = \frac{s_i^{\text{proxy}} - \min_j s_j^{\text{proxy}}}{\max_j s_j^{\text{proxy}} - \min_j s_j^{\text{proxy}}}.
\label{eq:proxy_score}
\end{equation} 

To assess whether the proxy is sufficiently confident, and given that PEX ultimately returns at most $\pi$ join paths, we probe whether the top-$\pi$ highest scores are well separated from the rest. To this end, we examine an indicator, namely the \emph{frontier gap} $\Delta_{\pi}$ in proxy scores between the $\pi$-th and the ($\pi$+1)-th ranked tables: 
\begin{equation}
\Delta_{\pi} = s^{\text{norm}}_{(\pi)} - s^{\text{norm}}_{(\pi+1)},
\end{equation}
Let $\tau$ denote a confidence threshold for ascertaining that the scores can be separated. If $\Delta_{\pi} \geq \tau$, PEX uses the normalized proxy scores directly. Otherwise, PEX issues a single batch LLM call to obtain semantic relevance scores. Note that if fewer than $\pi+1$ candidate tables exist, we directly invoke the LLM.

The LLM prompt (Appendix~\ref{appendix:prompts} at the end of this paper) encodes the prediction task, the base-table schema, the candidate-table schemas, and the feature descriptions generated by FDG. 
The LLM is queried once to obtain semantic relevance scores for all candidate tables in the range [0,100], subsequently normalized to [0,1]. 
This design evaluates all tables within a unified prompt context, promoting consistent scoring across candidates. 
If the number of candidate tables exceeds the LLM context window, we retain only 
candidate tables 
with the highest proxy scores and query the LLM on this subset; all remaining tables receive a semantic score of zero.
Ultimately, the output of this phase is a $SemScore(T_{i})$ assigned to each table $T_i$.

\subsubsection{Hybrid Path Scoring}
\label{subsec:hybrid-path-scoring}

\hippasus employs a breadth-first search that follows edges in both directions to explore join paths up to length $\ell$ starting from the base table. 
Crucially, while in the first phase we score individual tables, in the second phase we score complete paths using a hybrid scoring function that combines semantic table scores with statistical connection quality metrics defined below. 
The algorithm maintains a min-heap of the top-$\pi$ highest-scoring paths discovered during exploration. 
As paths are discovered, we score them immediately using the hybrid scoring function and update the top-$\pi$ heap incrementally, avoiding materialization of all possible paths. 
To prevent cycles, the algorithm checks whether each neighbor already appears in the current path before extension.

\vspace{2mm}
\noindent \textbf{Statistical Metrics.} 
We evaluate join feasibility between consecutive tables using three metrics computed from table metadata (requiring no join execution). 
For each hop from $T_i$ to $T_{i+1}$, where $T_i$ is the anchor and $T_{i+1}$ is the lookup table, let $\text{FK}_{T_i}$ denote the foreign key column in $T_i$, and $\text{JK}_{T_{i+1}}$ the join key column in $T_{i+1}$ that it references. 
Since traversal is bidirectional, $\text{JK}_{T_{i+1}}$ may be either a primary or a foreign key. 

\begin{itemize}[leftmargin=*]

\item \textit{Coverage} measures the fraction of anchor rows that will receive features:
\begin{equation}
\text{Cov}(T_i, T_{i+1}) = 1 - \text{NullRate}(\text{FK}_{T_i})
\end{equation}
Non-null foreign keys indicate successful matches, while null values result in missing features after the join. Low coverage leads to sparse feature columns in the augmented table.

\item \textit{Uniqueness} measures whether the join will cause row explosion:
\begin{equation}
\text{Uniq}(T_i, T_{i+1}) = \frac{|\{\text{distinct values in JK}_{T_{i+1}}\}|}{|T_{i+1}|}
\end{equation}
Values near 1 indicate that each key appears once in $T_{i+1}$, ensuring no fan-out. Lower values signify that multiple rows in $T_{i+1}$ share the same key, causing row explosion that can degrade both efficiency and model quality.

\item \textit{Size Ratio} reflects relative table sizes:
\begin{equation}
\text{SRatio}(T_i, T_{i+1}) = \frac{\min(|T_i|, |T_{i+1}|)}{\max(|T_i|, |T_{i+1}|)}
\end{equation}
This heuristic captures information diversity: very small lookup tables typically yield low-cardinality categorical features, while similarly sized tables tend to provide richer feature sets.

\end{itemize} 

\vspace{2mm}
\noindent \textbf{Scoring Function.} 
We score each join path by combining semantic and statistical signals. 
For a path $p = [T_{\text{base}}, T_1, \ldots, T_{\ell-1}]$, its \emph{hybrid score} is computed as: 
\begin{equation}
\text{Score}(p) = \frac{S_{\text{sem}} + S_{\text{stat}}}{2(\ell-1)} 
\end{equation} 
\noindent where $S_{\text{sem}}$ is the cumulative semantic score, $S_{\text{stat}}$ is the cumulative statistical score, and $2(\ell-1)$ 
corresponds to the number of score components (two per hop: one semantic, one statistical). 
The semantic score sums the semantic relevance scores assigned in the first phase (Section~\ref{subsec:semantic-table-scoring}) to all tables beyond the base: 
\begin{equation}
S_{\text{sem}} = \sum_{i=1}^{\ell-1} \textit{SemScore}(T_i)
\end{equation}
\noindent The statistical score aggregates connection quality for consecutive table pairs
based on the three aforementioned metrics: 
\begin{equation}
\begin{split}
S_{\text{stat}} = \sum_{i=0}^{\ell-2} \big[ & \alpha \cdot \textit{Cov}(T_i, T_{i+1})
+ \beta \cdot \textit{Uniq}(T_i, T_{i+1}) \\
& + \gamma \cdot \textit{SRatio}(T_i, T_{i+1}) \big]
\end{split}
\end{equation}
\noindent where $\alpha + \beta + \gamma = 1$ are weights (we use $\alpha = \beta = \gamma = 1/3$ by default). 
Normalizing by $2(\ell-1)$ ensures that scores lie in $[0,1]$ regardless of the path length $\ell$, preventing bias toward longer or shorter paths.

\subsection{Join Executor}
\label{sec:join-executor}  

The Join Executor (JEX) is responsible for materializing the join paths selected by the Path Explorer. JEX takes as input the selected paths $\mathcal{P}$ and the tables in the corpus, and produces $\pi$ augmented tables, one per path. Feature augmentation for supervised learning must satisfy two fundamental invariants: 
(1) \textit{row preservation}—the augmented table contains exactly $|T_{\text{base}}|$ rows, ensuring no training examples are duplicated or removed, and 
(2) \textit{distribution preservation}—the distribution of the target variable $y$ remains unchanged, preserving class frequencies (classification) or the empirical distribution (regression). 
Violating these invariants would alter the ML task itself, making the augmented dataset incompatible with the original supervised learning objective. 

To satisfy these invariants, existing systems~\cite{vldb20-arda, icde24-autofeat, icde25-featpilot} employ sequential binary left outer joins when materializing join paths. 
Given a join path $p = [T_{\text{base}}, T_1, T_2, \ldots, T_{\ell-1}]$
with join keys $(c_i^L, c_i^R)$ for each edge $(T_{i-1}, T_i)$, the binary approach sequentially materializes the path via left outer joins, i.e., $R_i \gets R_{i-1} \lojoin T_i'$, with $R_0 = T_{\text{base}}$, where $T_i'$ is the deduplicated version of $T_i$ on join key $c_i^R$. 
Deduplication ensures unique key values in $T_i'$, preventing row explosion, while left outer joins preserve all base table rows, thereby maintaining both invariants by construction. 
While correct, this sequential approach materializes all $\ell-2$ intermediate results at full size, potentially incurring  computational overhead for long paths. 

To address this inefficiency, we propose a strategy that leverages the classical Yannakakis algorithm~\cite{vldb81-yannakakis}, which efficiently computes acyclic joins through semi-join reductions. 
The key insight is to partition the join path into a prefix (the base table $T_{\text{base}}$) and a suffix ($[T_1, T_2, \ldots, T_{\ell-1}]$).
We apply Yannakakis-style semi-join reductions to the suffix tables, top-down from the base table, keeping in each table only the tuples whose key its parent references, and then left-join them into an intermediate result $S$.
We then left outer join this suffix result to the base table to obtain a path-specific joined table $R_p \gets T_{\text{base}} \lojoin S'$, where $S'$ is the deduplicated suffix on join key $c_1^R$. Since the base table enters only through this final left outer join, every base row is preserved, and rows whose join chain breaks receive nulls instead of being dropped. 
We refer to this approach as \emph{suffix-Yannakakis}.

To maintain the invariants when dealing with one-to-many or many-to-many join relationships, we perform deduplication, ensuring that each base table row joins with at most one tuple from each foreign table. 
Following prior work~\cite{vldb20-arda, icde24-autofeat, icde25-featpilot}, we deduplicate by selecting the first occurrence of a tuple among multiple tuples that share the same join key, converting relationships to one-to-one. 
Alternatively, an aggregation function (e.g., average, sum, count) can be applied. The selection of an appropriate aggregation function has been investigated in FeatAug~\cite{icde24-feataug}, which, however, operates in a different setting, involving a fixed pair of tables rather than a table corpus. Incorporating such a mechanism in JEX can be a future extension.  

Under deterministic table ordering and a fixed deduplication policy, suffix-Yannakakis produces the same rows and joined feature values as the sequential binary left-join strategy. 
In contrast, sequential binary left-join materialization may repeatedly construct full-size intermediate results along the path. 
When semi-join reductions are effective,
suffix-Yannakakis can substantially reduce both intermediate result size and execution cost.
As shown in prior work~\cite{vldb81-yannakakis, sigmod25-keyi, vldb25-stijn} and in our experiments, this effect is especially pronounced for acyclic join paths with selective suffix reductions.
We formalize both correctness and complexity of JEX in Appendix~\ref{appendix:jex-correctness-and-complexity} at the end of this paper.

After executing all $\pi$ selected join paths, JEX produces $\pi$ path-specific joined tables, each satisfying the invariants.
When multiple paths reach the same table, the same feature may appear in multiple such joined tables, each with a different null-value ratio, depending on the effectiveness of the join.  
In that case, we select the version with the lowest null ratio. 
Eventually, the consolidated table $T_{\text{cons}}$ is constructed by combining the base table with all selected features, maintaining $|T_{\text{cons}}| = |T_{\text{base}}|$ as consolidation performs only column operations. 
After consolidation, primary and foreign key columns are removed from the feature pool to avoid identifier leakage and spurious statistical prominence during feature selection.

\subsection{Feature Selector}
\label{sec:feature-selector} 

After consolidation, the output table $T_{\text{cons}}$ contains all base features plus additional features discovered through join path materialization. 
However, a large feature space may even degrade model performance, as it may lead to overfitting or it may contain redundant (e.g., correlated) or noisy features. 
\hippasus therefore employs a statistically grounded feature selection strategy with LLM-based semantic refinement to select a subset of $\kappa$ discovered features for inclusion in the final augmented table $T_{\text{aug}}$. 

FS operates only on the augmented features produced by JEX, while all base-table features are retained by construction.
Existing feature augmentation approaches~\cite{vldb20-arda, icde24-autofeat, icde25-featpilot} rely on statistical metrics to rank features based on their observed relationships with the target variable in training data. 
Following this principle, \hippasus computes feature statistics exclusively on the training split, ensuring that no test information influences feature selection.
Our primary statistical signal is \textit{Mutual Information} (MI), which measures the dependency between a feature $f$ and target $y$, quantifying how much knowing $f$ reduces uncertainty about $y$.
We use MI as the default ranking criterion because it applies uniformly across numerical and categorical features and effectively captures nonlinear dependency patterns.

Before invoking semantic analysis, FS consults the semantic-signal annotation produced by FDG. 
If semantic signal is unavailable, no LLM call is made, and the augmented features are ranked directly by MI in descending order, from which the top-$\kappa$ are selected. 
Otherwise, FS first applies a statistical prefilter when the number of augmented features is too large for the LLM context window. 
Specifically, it retains only the top-$K$ candidates in the MI ranking, where $K \gg \kappa$ (e.g., $K=100$), and passes them to the LLM together with their feature names, FDG-generated descriptions, and MI scores. 

To leverage the complementary strengths of empirical measurements and semantic analysis, \hippasus anchors the LLM on the MI ranking rather than asking it to rank features from scratch. 
We construct a prompt (Appendix~\ref{appendix:prompts} at the end of this paper) 
consisting of a system message that establishes the task context and a user message that provides the feature information. 
The prompt includes the task type (classification or regression), task description, and target column to establish context. 
For each candidate feature, we provide its rank in the MI ordering, its MI score, its name, and the description generated by FDG.
The features are presented pre-sorted by MI in descending order, and the LLM is instructed to make targeted adjustments to this ranking only when semantic evidence provides a clear reason to override the default MI order.
In particular, the prompt explicitly allows the LLM to demote high-MI artifacts, such as unique identifiers or leakage-prone fields, and to promote features whose descriptions reveal strong causal or domain relevance that MI-based ranking may underestimate~\cite{DBLP:journals/nca/VergaraE14}.
When uncertain, the LLM is instructed to preserve the MI order.
The LLM is prompted to rank all provided features with explicit constraints—each feature must appear exactly once in the output—and return the ranking as a JSON array ordered from most to least important. 
From the LLM's ranked output, we select the top-$\kappa$ features to produce the final augmented table $T_{\text{aug}}$, which is then used to train the downstream model. 

Thus, FS treats statistical evidence as the default ordering and semantic analysis as a constrained refinement mechanism.
This design makes the LLM a semantic auditor of the MI ranking rather than a free-form feature ranker, reducing sensitivity to model variation while preserving its ability to correct statistical artifacts.

\begin{table}[t]
\centering
\caption{\normalfont Dataset Statistics.}
\label{tab:datasets} 
\setlength{\tabcolsep}{1.7pt}
\makebox[\linewidth][c]{%
\begin{tabular}{lcccll}
\toprule
\textbf{Dataset} & \textbf{\#Tables} & \textbf{Hops} & \textbf{\#Features} & \textbf{\#Rows} & \textbf{Task / Domain} \\
\midrule
School & 15 & 1 & 702 & 1774 & C / School performance \\
Credit & 5 & 2 & 20 & 1000 & C / Credit risk \\
Eyemove & 6 & 2 & 24 & 7608 & C / Sentence relevance \\
Steel & 15 & 3 & 27 & 1941 & C / Steel fault \\
Jannis & 12 & 2 & 54 & 57580 & C / AutoML benchmark \\
Miniboone & 15 & 3 & 50 & 72998 & C / Neutrino events \\
Covertype & 12 & 2 & 54 & 423680 & C / Forest cover type \\
Bioresponse & 40 & 4 & 419 & 3434 & C / Biological response \\
Diabetes & 19 & 3 & 21 & 20000 & C / Diabetes prediction \\
Fraud & 16 & 3 & 27 & 20000 & C / Online fraud \\
Poverty & 25 & 3 & 55 & 3136 & R / Poverty level \\
Air & 23 & 3 & 134 & 11839 & R / Air quality \\
Northwind & 10 & 5 & 63 & 2155 & R / Order quantity \\
\bottomrule
\end{tabular}%
}
\end{table}

\section{Experimental Evaluation}
\label{sec:experiments} 

We evaluate \hippasus using a variety of publicly available real-world datasets and investigate: 
(i) how effective and efficient \hippasus is compared to state-of-the-art feature augmentation approaches, and how robust it is when the join graph is discovered rather than curated; 
(ii) how its adaptive use of statistical signals, SLMs, and LLMs affects predictive performance;
(iii) when semantic analysis improves feature augmentation and when cheaper alternatives suffice; 
(iv) how our join execution strategy impacts runtime; and 
(v) how sensitive \hippasus is to parameter choices, and how much of its gain comes from multi-hop paths. 

\begin{table*}[t]
\caption{\normalfont Effectiveness of \hippasus against baselines.}
\label{tab:end-to-end-effectiveness}
\centering
\small
\setlength{\tabcolsep}{2pt}
\begin{tabular}{l||c|c|c|c|c|c|c|c|c|c|c|c|c}
\hline
\multirow{2}{*}{\textbf{Variant}} & \textbf{School} & \textbf{Credit} & \textbf{Eyemove} & \textbf{Steel} & \textbf{Jannis} & \textbf{Miniboone} & \textbf{Covertype} & \textbf{Bioresponse} & \textbf{Diabetes} & \textbf{Fraud} & \textbf{Poverty} & \textbf{Air} & \textbf{Northwind} \\
& Acc. & Acc. & Acc. & Acc. & Acc. & Acc. & Acc. & Acc. & Acc. & F1 & MAE & RMSE & MAE \\
\hline
\hline
Base & 0.699 & 0.696 & 0.501 & 0.653 & 0.565 & 0.699 & 0.504 & 0.585 & 0.520 & 0.027 & 12028 & 13.899 & 14.035 \\
ARDA & \ul{0.802} & 0.696 & 0.539 & 0.717 & 0.565 & 0.879 & 0.595 & 0.642 & 0.520 & 0.375 & 12028 & 6.755 & 11.924 \\
AutoFeat & 0.702 & \ul{0.723} & 0.568 & \ul{0.768} & \ul{0.683} & 0.795 & \textbf{0.815} & \textbf{0.787} & 0.698 & 0.353 & \ul{11994} & \ul{6.617} & \ul{11.849} \\
FeatPilot & 0.794 & 0.701 & \textbf{0.588} & 0.754 & 0.565 & \textbf{0.891} & 0.642 & 0.692 & \ul{0.716} & \ul{0.538} & \textbf{11952} & 13.438 & 13.686 \\
\hippasus & \textbf{0.822} & \textbf{0.736} & \ul{0.577} & \textbf{0.775} & \textbf{0.713} & \ul{0.888} & \ul{0.767} & \ul{0.724} & \textbf{0.743} & \textbf{0.573} & 12035 & \textbf{5.769} & \textbf{11.797} \\
\hline
\end{tabular}
\end{table*}

\begin{table*}[t]
\caption{\normalfont Efficiency of \hippasus against baselines in seconds; for \hippasus, computation time + LLM wait time.}
\label{tab:end-to-end-efficiency}
\centering
\small
\setlength{\tabcolsep}{2pt}
\begin{tabular}{l||c|c|c|c|c|c|c|c|c|c|c|c|c}
\hline
\textbf{Variant} & \textbf{School} & \textbf{Credit} & \textbf{Eyemove} & \textbf{Steel} & \textbf{Jannis} & \textbf{Miniboone} & \textbf{Covertype} & \textbf{Bioresponse} & \textbf{Diabetes} & \textbf{Fraud} & \textbf{Poverty} & \textbf{Air} & \textbf{Northwind} \\
\hline
\hline
ARDA & 150 & 3 & 22 & 8 & 39 & 23 & 54 & 31 & 33 & 15 & 41 & 161 & 30 \\
AutoFeat & 8 & 2 & 3 & 3 & 15 & 17 & 69 & 32 & 7 & 8 & 5 & 26 & 6 \\
FeatPilot & 475 & 223 & 267 & 270 & 613 & 724 & 503 & 264 & 316 & 643 & 586 & 1103 & 340 \\
\hippasus & 3+18 & 0+8 & 1+6 & 0+2 & 7+0 & 8+0 & 61+4 & 1+0 & 2+2 & 3+7 & 1+11 & 3+9 & 1+17 \\
\hline
\end{tabular}
\end{table*}

\subsection{Experimental Setup}
\label{subsec:setup} 

\textbf{Datasets.} 
We evaluate \hippasus over 13 real-world datasets that have also been used by recent approaches~\cite{icde24-autofeat, icde25-featpilot}. 
Table~\ref{tab:datasets} shows for each dataset: the number of tables, 
the maximum number of joins needed to reach a table from the base table (hops), the total number of features, the number of rows in the base table, and the task/domain, i.e., classification (``C'') or regression (``R''), together with a short description of the prediction setting.
Specifically, we use \textbf{School} (predicting school performance based on student attributes on standardized tests); \textbf{Credit} (predicting whether an individual is a good or bad credit risk based on financial, demographic, and employment-related attributes); \textbf{Eyemove} (predicting sentence relevance using eye-tracking features); \textbf{Steel} (classifying fault type in steel plates using geometric, spatial, and luminosity-related features);
\textbf{Jannis} (classifying high-dimensional features from the ChaLearn AutoML benchmark);
\textbf{Miniboone} (distinguishing electron neutrino events from muon neutrino background using particle identification features); \textbf{Covertype} (predicting the presence of a specific forest cover type using normalized numerical and binary environmental features); \textbf{Bioresponse} (predicting the biological response of molecules from their molecular descriptors); \textbf{Diabetes} (predicting diabetes using features such as BMI and blood pressure); \textbf{Fraud} (predicting whether online transactions are fraudulent based on transaction and identity characteristics); \textbf{Poverty} (predicting poverty levels using socioeconomic features such as unemployment and education rates across U.S. states); 
\textbf{Air} (predicting the temperature in cities on given dates using features such as ozone and sulfur dioxide); 
and
\textbf{Northwind}~\cite{DBLP:journals/jisedu/DyerR15} (predicting order quantities in a relational sales database using features such as product categories, and supplier information).
Appendix~\ref{appendix:datasets} at the end of this paper describes how we preprocessed the datasets.

\vspace{2mm}
\noindent\textbf{SLM and LLM Models.}
For the SLM-based semantic proxy used in \hippasus, we employ the sentence-transformer \texttt{all-MiniLM-L6-v2}\footnote{\url{https://huggingface.co/sentence-transformers/all-MiniLM-L6-v2}}.
We test \hippasus with various LLMs including three commercial (\texttt{gpt-6-luna}, \texttt{gpt-6.1-sol}, \texttt{claude-sonnet-5.5}) and four open-weight models (\texttt{llama-4-scout}, \texttt{llama-4-maverick}, \texttt{ministral-14b}, \texttt{deepseek-v4-pro}, ranging from 14B to 1.6T parameters). 
Unless stated otherwise, all LLMs are queried with a temperature of $0.1$ to reduce output variance, and prompts are identical across models to ensure a fair comparison. 
All LLM calls are issued through OpenRouter\footnote{\url{https://openrouter.ai/}} using identical prompts across runs. 
The main results for \hippasus use \texttt{gpt-6-luna} as the LLM and \texttt{all-MiniLM-L6-v2} as the SLM proxy.

\vspace{2mm}
\noindent 
\textbf{Parameters.} Next, we explain the parameters used in the experiments; default values are in \textbf{bold}.  
We vary the maximum path length $\ell$ from $2$ to $\textbf{7}$. 
The number $\pi$ of join paths to explore takes values from $\{5,\textbf{10},20\}$. 
The cascade threshold $\tau$ is examined over $\{0.0005, \textbf{0.001}, 0.003, 0.005, 0.02, 0.05\}$. 
The number of extracted features $\kappa$ varies from $\{5, \textbf{10}, 15, 20\}$ per dataset.

\vspace{2mm}
\noindent\textbf{Evaluation Metrics.}
For classification tasks, we report \emph{accuracy} for balanced datasets and \emph{F1-score} for imbalanced datasets. 
For regression tasks, we report \emph{Mean Absolute Error} (MAE) or \emph{Root Mean Squared Error} (RMSE), following prior work. 
Efficiency is measured as the total \emph{feature augmentation time} (in seconds), including path exploration, join execution, and feature selection, on an Apple M4 Max laptop with 36\,GB of memory. 
To quantify the cost of semantic analysis, we additionally report the number of input and output tokens consumed by LLM calls. 
We exclude downstream model training time, as it is orthogonal to feature augmentation. 
All reported results are averages over five runs. 
We use AutoGluon~\cite{arxiv20-autogluon-tabular} to automatically train multiple models (random forests, gradient boosting, and extra trees) and ensemble them to create the final predictor. 
In the results, values in \textbf{bold} indicate the best performance per dataset;  \underline{underlined} values indicate second-best performance. 

\vspace{2mm}
\noindent \textbf{Baselines.} 
We compare \hippasus with three state-of-the-art feature augmentation methods: \textbf{ARDA}~\cite{vldb20-arda}, \textbf{AutoFeat}~\cite{icde24-autofeat}, and \textbf{FeatPilot}~\cite{icde25-featpilot} (see Section~\ref{sec:related-work}). 
For AutoFeat and FeatPilot, we use the publicly released implementations from the authors. 
We have implemented ARDA following the algorithmic details in~\cite{vldb20-arda}, as the original code was not available. 
For all methods, we use the default parameters from the respective papers. 
We supply ARDA with star schemata, since it is limited to this setting. 
FeatPilot runs unmodified: it trains its LSTM on its own exploration of the same tables, using only the training split shared by all methods. 
We do not include AutoFeature~\cite{icde22-autofeature} in our comparison, as recent work~\cite{icde24-autofeat, icde25-featpilot} has shown both AutoFeat and FeatPilot outperform it in effectiveness and efficiency.
Moreover, we denote by \textbf{Base} a method that uses the original base table without any augmented features. 
This serves as a reference to evaluate the impact of feature augmentation. 
All baselines use the same ML training and evaluation pipeline through AutoGluon~\cite{arxiv20-autogluon-tabular} to ensure a fair comparison. 

\subsection{Comparison with Baselines}  
\label{subsec:end-to-end} 

Using the default parameters, we evaluate end-to-end performance, which reflects the quality of the final downstream model after the complete feature augmentation pipeline. 

\noindent \textbf{Effectiveness.} 
As listed in Table~\ref{tab:end-to-end-effectiveness}, \hippasus achieves the best performance on 8 out of 13 datasets and the second best on 4, with average improvements of 15.7\% over ARDA, 8.1\% over AutoFeat, and 10.8\% over FeatPilot. On the remaining dataset, \textit{Poverty}, no method improves the MAE of Base by more than 0.7\%, suggesting that the candidate tables add little signal for this target. 
We attribute this to several limitations or weaknesses of existing approaches, which \hippasus overcomes via its hybrid (statistics- and semantics-driven) method.
ARDA's 1-hop exploration strategy prevents it from discovering distant features, leading to poor performance on multi-hop schemas such as \textit{Eyemove}, \textit{Fraud}, and \textit{Diabetes}, while excelling on \textit{School}'s star schema, where informative features are within one hop. Section~\ref{subsec:effect-of-parameters} measures how much of \hippasus's gain comes from multi-hop features. 
AutoFeat ranks join paths using cheaper statistical metrics (relevancy and redundancy scores), which fail to effectively address search space complexity, resulting in suboptimal path selection that misses semantically meaningful relationships. 
FeatPilot trains LSTM models to predict join path quality and employs embeddings for feature clustering, but this added complexity yields marginal benefits, suggesting that embeddings alone are insufficient for capturing task-specific semantic relevance in path exploration. 
Finally, the Base method shows poor performance, demonstrating the clear value of feature augmentation.

\noindent \textbf{Efficiency.} 
As shown in Table~\ref{tab:end-to-end-efficiency}, \hippasus maintains competitive efficiency, with substantial speedups over complex baselines (ARDA and FeatPilot). For \hippasus, each cell shows the time of its own computation plus the time spent waiting for LLM replies, which depends on the LLM provider rather than on \hippasus; its own computation is the fastest among all methods on every dataset except Covertype. Its computation grows with the number of base-table rows that joins and feature selection process: on \textit{Covertype}, with 424K rows, join execution takes 35\,s and feature selection 15\,s of its 61\,s, while every other dataset takes at most 8\,s (Appendix~\ref{appendix:time-breakdown} at the end of this paper). 
Note that the execution time of the Base method is zero, since it does not perform any feature augmentation. 
\hippasus is faster than FeatPilot on all datasets, with a median speedup of about $64\times$, as FeatPilot trains LSTM models during the augmentation process to predict join path quality. 
Against ARDA, \hippasus achieves a median speedup of about $3.4\times$, since ARDA exhaustively materializes all 1-hop joins and incorporates the ML model into its feature discovery process. 
Relative to AutoFeat, which uses lightweight statistical heuristics and no LLM, \hippasus is on par despite its LLM calls: it is faster on 7 of the 13 datasets and takes 13.5\,s per dataset on average against 15.5\,s, while being more effective, as discussed earlier. 
\hippasus's efficiency stems from its decoupled architecture: the path exploration stage ranks candidate paths using adaptive semantic scoring without materializing joins, then the join execution stage selectively executes only top-ranked paths using the suffix-Yannakakis algorithm.

\begin{table*}[t]
\centering
\caption{\normalfont Effectiveness on discovered join graphs (TO: timeout; every run exceeded the 30-minute limit). Last row: relative change of \hippasus from the curated graphs (Table~\ref{tab:end-to-end-effectiveness}) to the discovered ones (positive: better).}
\label{tab:discovered-join-graphs}
\small
\setlength{\tabcolsep}{1.6pt}
\begin{tabular}{l||c|c|c|c|c|c|c|c|c|c|c|c|c}
\hline
\multirow{2}{*}{\textbf{Method}} & \textbf{School} & \textbf{Credit} & \textbf{Eyemove} & \textbf{Steel} & \textbf{Jannis} & \textbf{Miniboone} & \textbf{Covertype} & \textbf{Bioresponse} & \textbf{Diabetes} & \textbf{Fraud} & \textbf{Poverty} & \textbf{Air} & \textbf{Northwind} \\
& Acc. & Acc. & Acc. & Acc. & Acc. & Acc. & Acc. & Acc. & Acc. & F1 & MAE & RMSE & MAE \\
\hline
\hline
ARDA & 0.787 & 0.696 & 0.530 & 0.715 & 0.566 & \underline{0.879} & 0.585 & 0.642 & 0.520 & 0.345 & 12028 & \underline{6.753} & 13.747 \\
AutoFeat & TO & 0.694 & 0.567 & \textbf{0.766} & \underline{0.672} & 0.796 & \textbf{0.815} & \underline{0.681} & \textbf{0.700} & 0.398 & \textbf{11691} & 9.321 & \underline{11.881} \\
FeatPilot & \underline{0.788} & \underline{0.698} & \underline{0.582} & 0.741 & 0.563 & 0.878 & 0.625 & 0.680 & 0.604 & \underline{0.614} & \underline{11882} & 12.129 & 13.853 \\
\hippasus & \textbf{0.811} & \textbf{0.736} & \textbf{0.587} & \underline{0.744} & \textbf{0.764} & \textbf{0.887} & \underline{0.770} & \textbf{0.711} & \underline{0.699} & \textbf{0.644} & 12014 & \textbf{5.874} & \textbf{11.766} \\
\hline
\makecell[l]{\hippasus{} vs.\\curated (\%)} & $-$1.3 & 0.0 & +1.7 & $-$4.0 & +7.2 & $-$0.1 & +0.4 & $-$1.8 & $-$5.9 & +12.4 & +0.2 & $-$1.8 & +0.3 \\
\hline
\end{tabular}
\end{table*}

\begin{table*}[t]
\centering
\caption{\normalfont Effect of semantic analysis.}
\label{tab:system-ablation}
\small
\setlength{\tabcolsep}{1.6pt}
\begin{tabular}{l||c|c|c|c|c|c|c|c|c|c|c|c|c}
\hline
\multirow{2}{*}{\textbf{Variant}} & \textbf{School} & \textbf{Credit} & \textbf{Eyemove} & \textbf{Steel} & \textbf{Jannis} & \textbf{Miniboone} & \textbf{Covertype} & \textbf{Bioresponse} & \textbf{Diabetes} & \textbf{Fraud} & \textbf{Poverty} & \textbf{Air} & \textbf{Northwind} \\
& Acc. & Acc. & Acc. & Acc. & Acc. & Acc. & Acc. & Acc. & Acc. & F1 & MAE & RMSE & MAE \\
\hline
\hline
$\hippasus_{\textit{stats}}$ & 0.700 & \textbf{0.745} & \textbf{0.656} & \textbf{0.777} & \textbf{0.713} & \textbf{0.888} & \underline{0.845} & \textbf{0.724} & \underline{0.741} & \textbf{0.632} & 12156 & 8.233 & 12.138 \\
$\hippasus_{\textit{SLM}}$ & 0.820 & 0.702 & \underline{0.640} & \underline{0.775} & \textbf{0.713} & \textbf{0.888} & 0.767 & \textbf{0.724} & \underline{0.741} & 0.565 & \textbf{11905} & \underline{5.782} & \underline{11.810} \\
$\hippasus_{\textit{LLM}}$ & \textbf{0.828} & \underline{0.736} & 0.577 & 0.760 & \textbf{0.713} & \textbf{0.888} & \textbf{0.963} & \textbf{0.724} & 0.719 & \underline{0.573} & 12071 & 6.048 & \textbf{11.797} \\
$\hippasus$ & \underline{0.822} & \underline{0.736} & 0.577 & \underline{0.775} & \textbf{0.713} & \textbf{0.888} & 0.767 & \textbf{0.724} & \textbf{0.743} & \underline{0.573} & \underline{12035} & \textbf{5.769} & \textbf{11.797} \\
\hline
\end{tabular}
\end{table*}

\vspace{2mm}
\noindent\textbf{Robustness to discovered join graphs.}
The join graphs of our benchmarks are curated: their key--foreign key relationships are given as input. As Section~\ref{sec:problem-statement} allows, they can also come from discovery tools, whose output may be wrong. To test robustness to such errors, we replace each curated join graph with one discovered from the tables alone, by a foreign-key detector that we implemented from established rules~\cite{pvldb14-chen-fk, webdb09-rostin}: a column $a$ joins a column $b$ of another table if $b$ is key-like (at least 95\% of its values distinct), both hold integers or strings, and $a$ has at least five distinct values, of which at least half occur in $b$ (containment) and cover at least half of $b$'s values (coverage); each column joins only its best-matching key. We fixed these thresholds before any run. The discovered graphs add spurious edges on ten datasets, from 1 on \textit{Fraud} to 194 of the 206 edges on \textit{School}, where they raise the candidate paths from 15 to 1,153; on four datasets, they also cut correct access to tables, e.g., only 8 of the 19 tables of \textit{Diabetes} remain reachable through correct joins. As FeatPilot requires the two columns of a join to have the same name, we give the two columns of each discovered join a common key name, as FeatPilot's own converter does for the curated graphs. With the tables, splits, seeds, and trainer of Table~\ref{tab:end-to-end-effectiveness}, \hippasus is the best method on nine datasets and second on three (Table~\ref{tab:discovered-join-graphs}). It stays within 2\% of its results on the curated graphs on nine datasets and even improves on \textit{Jannis} and \textit{Fraud}, by 7\% and 12\%. On \textit{School}, despite the 1,153 candidate paths, it still selects on average 8.8 of the 10 features it selects on the curated graph, while AutoFeat exceeds the 30-minute limit on every run. It drops by more than 2\% only on \textit{Steel} (4\%) and \textit{Diabetes} (6\%), and both drops trace back to discovery errors: spurious edges change three of its ten features on \textit{Steel}, and most tables of \textit{Diabetes} are no longer reachable through correct joins. The baselines suffer larger drops: AutoFeat loses 13\% on \textit{Bioresponse} and its RMSE on \textit{Air} rises by 41\%, ARDA's MAE on \textit{Northwind} rises by 15\%, and FeatPilot loses 16\% on \textit{Diabetes}.

\subsection{Ablation Study}
\label{subsec:ablation-study} 

To quantify the contribution of \hippasus's key design choices, we perform an ablation study that evaluates the effect of each core component on final prediction accuracy and efficiency. 
We study the role of semantic analysis for path exploration and feature selection, the role of feature descriptions in providing semantic signals for path exploration and feature selection, and the efficiency of the suffix-Yannakakis join algorithm relative to binary joins. 

\vspace{2mm}
\noindent \textbf{Semantic Analysis.} 
We compare four variants of \hippasus that differ in how semantic signals are used in both the Path Explorer (PEX) and the Feature Selector (FS).
$\hippasus_{\textit{stats}}$ disables semantic analysis entirely: in PEX, the semantic path score is set to zero ($S_{\text{sem}}{=}0$), and in FS, augmented features are ranked purely by statistics.
$\hippasus_{\textit{SLM}}$ uses SLM-based semantic analysis throughout: in PEX, semantic table relevance is derived from cosine similarity between sentence embeddings, while in FS, the statistical and semantic rankings are merged into a unified ranking using Borda count~\cite{wang2024survey}. 
$\hippasus_{\textit{LLM}}$ uses LLM-based semantic analysis in both modules: PEX always invokes the LLM oracle for semantic table scoring, and FS always applies LLM-based semantic reranking. 
Finally, our full system \hippasus combines these signals adaptively: in PEX, it uses the frontier-gap cascade policy with $\tau{=}0.001$ to decide whether the SLM proxy suffices or whether the LLM oracle is needed, while in FS it applies semantic reranking only when semantic signal is available.

Table~\ref{tab:system-ablation} reports the effectiveness of the four variants. 
Semantic analysis pays where the useful columns are named for what they measure: without it, $\hippasus_{\textit{stats}}$ stays at the accuracy of Base on \textit{School} (0.700 against 0.822), and its RMSE on \textit{Air} is 43\% higher than that of \hippasus. Where these columns are codes or positions, as on \textit{Eyemove} and \textit{Fraud}, statistics alone do better. \hippasus has the best average rank of the four variants: it is best or second best on eight of the ten datasets with semantic signal, against at most six for any fixed variant, and stays within about 1\% of the best variant on seven of them, while consuming about half the LLM input tokens of $\hippasus_{\textit{LLM}}$ (2.9K versus 6.2K per run).
Finally, on datasets with no semantic signal (\textit{Jannis}, \textit{Miniboone}, and \textit{Bioresponse}), all four variants reduce to the same statistics-driven behavior.

\vspace{2mm}
\noindent\textbf{Feature Description Generation.}
We compare two configurations: $\hippasus_{\textit{w/ FDG}}$ with generated descriptions and semantic-signal annotation and $\hippasus_{\textit{w/o FDG}}$ without them.  
FDG plays two roles. First, it decides whether semantic analysis is used at all: on \textit{Jannis}, \textit{Miniboone}, and \textit{Bioresponse}, whose original names and generated descriptions remain uninformative, FDG marks semantic signal as unavailable, so \hippasus makes no LLM call and relies on statistical signals alone. Second, its descriptions improve or preserve accuracy when the column names are informative, such as credit\_amount or BMI, so that each feature receives its own description, as on \textit{Credit}, \textit{Diabetes}, \textit{Poverty}, and \textit{Northwind}; when the names are opaque codes, such as V14, or many features share one description, the descriptions add little beyond the statistical signals (Appendix~\ref{appendix:fdg} at the end of this paper).

\begin{figure}[t]
    \centering
    \includegraphics[width=\columnwidth]{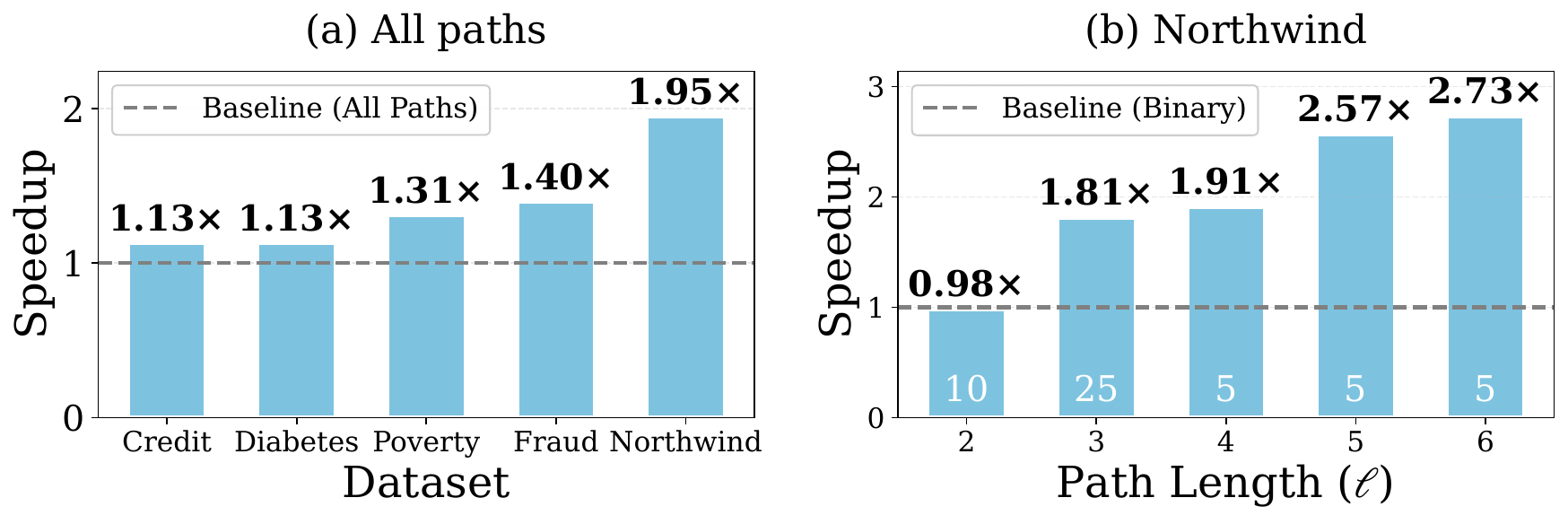}
    \caption{\normalfont Performance comparison of Suffix-Yannakakis approach. 
    (a) Speedup over all paths across different datasets. 
    (b) Speedup over binary join by path length on the \textit{Northwind} dataset.}
    \label{fig:speedup_comparison}
\end{figure} 

\begin{table}[t]
\centering
\caption{\normalfont Cascade routing and LLM cost. 
For proxy-routed datasets, we also report token savings relative to $\hippasus_{\textit{LLM}}$.}
\label{tab:cascade-routing}
\small
\setlength{\tabcolsep}{2.0pt}
\begin{tabular}{lcrrrr}
\hline
\textbf{Dataset} & \textbf{Routing} & \textbf{Input} & \textbf{Input} & \textbf{Output} & \textbf{Output} \\
 &  & \textbf{Tokens} & \textbf{Savings} & \textbf{Tokens} & \textbf{Savings} \\
\hline
School & proxy & 6264 & 16647 & 2184 & 800 \\
Credit & oracle & 2399 & -- & 404 & -- \\
Eyemove & oracle & 2516 & -- & 573 & -- \\
Steel & proxy & 911 & 2426 & 131 & 794 \\
Jannis & stats only & 0 & -- & 0 & -- \\
Miniboone & stats only & 0 & -- & 0 & -- \\
Covertype & proxy & 1752 & 2879 & 351 & 546 \\
Bioresponse & stats only & 0 & -- & 0 & -- \\
Diabetes & proxy & 1093 & 1902 & 140 & 210 \\
Fraud & oracle & 3542 & -- & 592 & -- \\
Poverty & proxy & 1433 & 3300 & 999 & 582 \\
Air & proxy & 2613 & 6319 & 573 & 937 \\
Northwind & oracle & 6315 & -- & 1628 & -- \\
\hline
\end{tabular}
\end{table}

\begin{table}[t]
\centering
\caption{\normalfont Effect of the cascade threshold $\tau$, relative to always using the LLM.}
\label{tab:cascade-threshold}
\small
\setlength{\tabcolsep}{4pt}
\begin{tabular}{rccc}
\hline
\textbf{$\tau$} & \textbf{Oracle calls} & \textbf{Input tokens} & \textbf{Accuracy change} \\
\hline
0.0005 & 30\% & 42\% & $-$1.18\% \\
\textbf{0.001} & \textbf{40\%} & \textbf{46\%} & \textbf{$-$1.07\%} \\
0.003 & 50\% & 56\% & $-$1.53\% \\
0.005 & 70\% & 87\% & $-$1.66\% \\
0.02 & 90\% & 95\% & $-$2.04\% \\
0.05 & 100\% & 100\% & $\phantom{-}$0.00\% \\
\hline
\end{tabular}
\end{table}

\vspace{2mm}
\noindent \textbf{Join Execution Strategy.} 
To evaluate the efficiency gains of our suffix-Yannakakis join strategy compared to the traditional binary approach, we compare the two strategies across representative datasets. 
Figure~\ref{fig:speedup_comparison}(a) presents the average speedup across datasets, showing that suffix-Yannakakis achieves consistent gains ranging from 1.13$\times$ on \textit{Credit} to 1.95$\times$ on \textit{Northwind}, demonstrating that multi-way join execution with semi-join reductions outperforms binary joins even on moderately complex schemas. 
Figure~\ref{fig:speedup_comparison}(b) shows speedup per path length on \textit{Northwind}. At path length $\ell=2$, both strategies perform identically since suffix-Yannakakis reduces to binary joins for single-hop paths, but from path length $\ell=3$ onward, semi-join reductions eliminate non-result tuples early, with speedup growing to 2.73$\times$ at $\ell=6$. 
This pattern is particularly pronounced on schemas with 1-to-many relationships, where suffix-Yannakakis prunes large intermediate results before materialization. 

\subsection{Adaptive Semantic Analysis}   
\label{subsec:adaptive-semantic-reasoning} 

We next examine the role of adaptive semantic analysis in \hippasus from three complementary perspectives. 

\begin{table*}[t]
\centering
\caption{\normalfont Effect of LLM model in \hippasus.}
\label{tab:llm-model-ablation-icde}
\footnotesize
\setlength{\tabcolsep}{2.8pt}
\begin{tabular}{l||c|c|c|c|c|c|c|c|c|c|c|c|c}
\hline
\multirow{2}{*}{\textbf{LLM Model}} & \textbf{School} & \textbf{Credit} & \textbf{Eyemove} & \textbf{Steel} & \textbf{Jannis} & \textbf{Miniboone} & \textbf{Covertype} & \textbf{Bioresponse} & \textbf{Diabetes} & \textbf{Fraud} & \textbf{Poverty} & \textbf{Air} & \textbf{Northwind} \\
& Acc. & Acc. & Acc. & Acc. & Acc. & Acc. & Acc. & Acc. & Acc. & F1 & MAE & RMSE & MAE \\
\hline
\hline
Llama-4-Scout & 0.700 & \textbf{0.746} & 0.590 & \textbf{0.775} & \textbf{0.713} & \textbf{0.888} & \textbf{0.767} & \textbf{0.724} & \underline{0.743} & 0.623 & 12067 & \underline{5.878} & 12.168 \\
Llama-4-Maverick & 0.704 & \underline{0.740} & \textbf{0.670} & \textbf{0.775} & \textbf{0.713} & \textbf{0.888} & \textbf{0.767} & \textbf{0.724} & \textbf{0.745} & \textbf{0.644} & 12151 & \textbf{5.769} & 11.914 \\
Ministral-14B & 0.815 & \textbf{0.746} & 0.589 & \textbf{0.775} & \textbf{0.713} & \textbf{0.888} & \underline{0.765} & \textbf{0.724} & \underline{0.743} & 0.632 & \underline{11978} & 6.152 & 12.068 \\
DeepSeek-V4-Pro & 0.700 & 0.735 & 0.562 & \textbf{0.775} & \textbf{0.713} & \textbf{0.888} & \textbf{0.767} & \textbf{0.724} & \underline{0.743} & 0.626 & \textbf{11950} & \textbf{5.769} & 12.140 \\
GPT-6-Luna & 0.822 & 0.736 & 0.577 & \textbf{0.775} & \textbf{0.713} & \textbf{0.888} & \textbf{0.767} & \textbf{0.724} & \underline{0.743} & 0.573 & 12035 & \textbf{5.769} & \textbf{11.797} \\
GPT-6.1-Sol & \textbf{0.828} & 0.736 & 0.607 & \textbf{0.775} & \textbf{0.713} & \textbf{0.888} & \textbf{0.767} & \textbf{0.724} & \underline{0.743} & \underline{0.635} & 12085 & \textbf{5.769} & \underline{11.800} \\
Claude-Sonnet-5.5 & \underline{0.823} & 0.736 & \underline{0.635} & \textbf{0.775} & \textbf{0.713} & \textbf{0.888} & \textbf{0.767} & \textbf{0.724} & \underline{0.743} & 0.536 & 12156 & \textbf{5.769} & 11.858 \\
\hline
\end{tabular}
\end{table*}

\vspace{2mm}
\noindent \textbf{Cascade routing policy and LLM token usage.} 
First, we study how the cascade routes datasets between statistics-only processing, SLM-based proxy reasoning, and LLM-based oracle reasoning, and quantify the resulting LLM token usage (Table~\ref{tab:cascade-routing}). \textit{Jannis}, \textit{Miniboone}, and \textit{Bioresponse} provide no semantic signal and therefore reduce to statistics-only behavior, incurring zero LLM cost. The cascade sends six of the ten datasets with semantic signal to the SLM proxy; it invokes the oracle on \textit{Credit}, \textit{Eyemove}, and \textit{Northwind}, whose join graphs are too small to yield a meaningful frontier gap, and on \textit{Fraud}, where the SLM proxy is uncertain.
For proxy-routed datasets, the table also reports token savings relative to $\hippasus_{\textit{LLM}}$, which always invokes the oracle in PEX. These savings are substantial: adaptive routing saves 16,647 input tokens on \textit{School}, 6,319 on \textit{Air}, and 3,300 on \textit{Poverty}, cutting the input tokens of $\hippasus_{\textit{LLM}}$ by 62--73\% on the proxy-routed datasets.
Token usage also varies strongly with schema and feature-set size, from 2,399 input tokens on \textit{Credit} to 6,315 on \textit{Northwind} among oracle-routed datasets, which shows that the cost of semantic analysis depends strongly on dataset complexity.

\vspace{2mm}
\noindent  
\textbf{Cascade threshold.} 
We evaluate the sensitivity of \hippasus to the cascade threshold $\tau$ over $\tau \in \{0.0005, 0.001, 0.003, 0.005, 0.02, 0.05\}$. Since $\tau$ only decides, per dataset, whether the cascade keeps the SLM proxy or invokes the LLM oracle, the outcome at each threshold follows from our runs of these two routes. 
Table~\ref{tab:cascade-threshold} reports, relative to always invoking the LLM oracle, the oracle invocation rate, the input tokens, and the average accuracy change over the ten datasets with semantic signal.
At $\tau{=}0.001$, the cascade invokes the oracle on four of the ten datasets and spends 46\% of the input tokens of always invoking it, while staying within 1.1\% of its accuracy on average, the best accuracy among all thresholds that save tokens. A smaller threshold also routes \textit{Fraud} to the SLM proxy and loses accuracy, while larger thresholds invoke the oracle on more datasets and raise the cost; only invoking it on every dataset ($\tau{=}0.05$) recovers the remaining 1.1\%, at more than twice the tokens. 
We therefore select $\tau{=}0.001$ as the default operating point, as it provides the best observed effectiveness--cost tradeoff.

\vspace{2mm}

\noindent \textbf{LLM model sensitivity.} To assess the impact of the LLM backend on effectiveness, we evaluate seven models: three commercial (\texttt{gpt-6-luna}, \texttt{gpt-6.1-sol}, \texttt{claude-sonnet-5.5}) and four open-weight (\texttt{llama-4-scout}, \texttt{llama-4-maverick}, \texttt{ministral-14b}, \texttt{deepseek-v4-pro}); all other parameters are set to the default values. Table~\ref{tab:llm-model-ablation-icde} lists the effectiveness results across all datasets. On average, \hippasus is robust to the choice of LLM: no single model dominates, and different models achieve the best result on different datasets. GPT-6.1-Sol and Llama-4-Maverick are the most consistent, ranking among the two best on seven of the ten datasets with semantic signal, while the default GPT-6-Luna attains the best result on four of them. Within a family, larger models yield modest gains (GPT-6.1-Sol over GPT-6-Luna, Llama-4-Maverick over Llama-4-Scout), but at a much higher cost: GPT-6.1-Sol costs 15$\times$ more per run than GPT-6-Luna. Open models are competitive with commercial ones: Llama-4-Maverick ranks second overall and Ministral-14B is on par with the commercial models, offering viable privacy-preserving options for organizations with data sensitivity constraints. The models differ far more in cost than in effectiveness (Table~\ref{tab:llm-model-token-usage-icde}): GPT-6-Luna is the least expensive, and no other model is both cheaper and more accurate, which is why we use it by default. On datasets with no semantic signal (\textit{Jannis}, \textit{Miniboone}, and \textit{Bioresponse}), all models yield identical effectiveness.

\begin{table}[t]
\centering
\caption{\normalfont Average token usage and cost per run.}
\label{tab:llm-model-token-usage-icde}
\begin{tabular}{lrrr}
\hline
\textbf{LLM Model} & \textbf{Input} & \textbf{Output} & \textbf{\$/1K runs} \\
\hline
Llama-4-Scout & 2,839 & 866 & 0.77 \\
Llama-4-Maverick & 2,874 & 1,260 & 2.20 \\
Ministral-14B & 3,339 & 953 & 0.85 \\
DeepSeek-V4-Pro & 2,954 & 1,174 & 3.55 \\
GPT-6-Luna & 2,884 & 757 & 0.66 \\
GPT-6.1-Sol & 2,868 & 453 & 10.08 \\
Claude-Sonnet-5.5 & 4,336 & 718 & 15.86 \\
\hline
\end{tabular}
\end{table}

Table~\ref{tab:llm-model-token-usage-icde} reports the average input and output tokens consumed per model, together with the cost per 1,000 runs. For every model, token usage is highest on feature-rich datasets such as \textit{School}, at two to three times its average, as more candidate features reach semantic reranking.

\subsection{Effect of Parameters}
\label{subsec:effect-of-parameters}

To understand how \hippasus's key hyperparameters affect effectiveness, we conduct sensitivity analysis following the natural dependency of \hippasus's pipeline: 
(i)   maximum join path length $\ell$, 
(ii)  number $\pi$ of join paths to materialize, and 
(iii) number $\kappa$ of features to finally select. 

\begin{table*}[t]
\centering
\caption{\normalfont Effect of the maximum join path length.}
\label{tab:sensitivity-max-path-depth}
\small
\setlength{\tabcolsep}{2pt}
\begin{tabular}{l||c|c|c|c|c|c|c|c|c|c|c|c|c}
\hline
\multirow{2}{*}{\textbf{Max $\ell$}} & \textbf{School} & \textbf{Credit} & \textbf{Eyemove} & \textbf{Steel} & \textbf{Jannis} & \textbf{Miniboone} & \textbf{Covertype} & \textbf{Bioresponse} & \textbf{Diabetes} & \textbf{Fraud} & \textbf{Poverty} & \textbf{Air} & \textbf{Northwind} \\
& Acc. & Acc. & Acc. & Acc. & Acc. & Acc. & Acc. & Acc. & Acc. & F1 & MAE & RMSE & MAE \\
\hline
\hline
2 & \textbf{0.822} & \underline{0.668} & \underline{0.587} & 0.715 & \underline{0.565} & \underline{0.879} & 0.641 & \underline{0.690} & 0.645 & 0.443 & \underline{12094} & 6.520 & 11.824 \\
3 & \textbf{0.822} & \textbf{0.736} & \textbf{0.668} & \underline{0.773} & \textbf{0.713} & \textbf{0.888} & \textbf{0.773} & \textbf{0.724} & 0.739 & \textbf{0.602} & \textbf{12035} & \underline{5.862} & 11.857 \\
4 & \textbf{0.822} & \textbf{0.736} & \underline{0.587} & \textbf{0.775} & \textbf{0.713} & \textbf{0.888} & \underline{0.767} & \textbf{0.724} & \textbf{0.745} & \underline{0.584} & \textbf{12035} & \underline{5.862} & 11.857 \\
5 & \textbf{0.822} & \textbf{0.736} & 0.577 & \textbf{0.775} & \textbf{0.713} & \textbf{0.888} & \underline{0.767} & \textbf{0.724} & \underline{0.743} & 0.580 & \textbf{12035} & \textbf{5.769} & \underline{11.819} \\
6 & \textbf{0.822} & \textbf{0.736} & 0.577 & \textbf{0.775} & \textbf{0.713} & \textbf{0.888} & \underline{0.767} & \textbf{0.724} & \underline{0.743} & 0.573 & \textbf{12035} & \textbf{5.769} & \textbf{11.797} \\
7 & \textbf{0.822} & \textbf{0.736} & 0.577 & \textbf{0.775} & \textbf{0.713} & \textbf{0.888} & \underline{0.767} & \textbf{0.724} & \underline{0.743} & 0.573 & \textbf{12035} & \textbf{5.769} & \textbf{11.797} \\
\hline
\end{tabular}
\end{table*}

\begin{table*}[t]
\centering
\caption{\normalfont Effect of the number of materialized join paths.}
\label{tab:sensitivity-num-join-paths}
\small
\setlength{\tabcolsep}{2pt}
\begin{tabular}{l||c|c|c|c|c|c|c|c|c|c|c|c|c}
\hline
\multirow{2}{*}{\textbf{Paths $\pi$}} & \textbf{School} & \textbf{Credit} & \textbf{Eyemove} & \textbf{Steel} & \textbf{Jannis} & \textbf{Miniboone} & \textbf{Covertype} & \textbf{Bioresponse} & \textbf{Diabetes} & \textbf{Fraud} & \textbf{Poverty} & \textbf{Air} & \textbf{Northwind} \\
& Acc. & Acc. & Acc. & Acc. & Acc. & Acc. & Acc. & Acc. & Acc. & F1 & MAE & RMSE & MAE \\
\hline
\hline
5 & \textbf{0.825} & \underline{0.728} & \underline{0.622} & 0.760 & 0.627 & \underline{0.890} & \underline{0.767} & 0.712 & 0.701 & 0.542 & \textbf{11784} & 6.010 & \underline{11.901} \\
10 & \underline{0.822} & \textbf{0.736} & 0.577 & \textbf{0.775} & \underline{0.713} & 0.888 & \underline{0.767} & \underline{0.724} & \textbf{0.743} & \underline{0.573} & 12035 & \textbf{5.769} & \textbf{11.797} \\
20 & 0.816 & \textbf{0.736} & \textbf{0.636} & \underline{0.774} & \textbf{0.769} & \textbf{0.896} & \textbf{0.931} & \textbf{0.728} & \underline{0.722} & \textbf{0.601} & \underline{12005} & \underline{5.862} & \textbf{11.797} \\
\hline
\end{tabular}
\end{table*}

\begin{table}[t]
\centering
\caption{\normalfont Contribution of multi-hop features: share of the selected features reached through two or more joins, and the gain over Base attributed to one-hop and to multi-hop features (Shapley split; in each dataset's metric, positive is better, larger in bold).}
\label{tab:multihop}
\small
\setlength{\tabcolsep}{3pt}
\begin{tabular}{llccc}
\hline
\textbf{Dataset} & \textbf{Metric} & \textbf{2+ hops} & \textbf{Gain, 1 hop} & \textbf{Gain, 2+ hops} \\
\hline
School & Acc. & 0\% & \textbf{+0.122} & +0.001 \\
Credit & Acc. & 74\% & +0.000 & \textbf{+0.040} \\
Eyemove & Acc. & 48\% & +0.031 & \textbf{+0.044} \\
Steel & Acc. & 40\% & +0.045 & \textbf{+0.077} \\
Jannis & Acc. & 94\% & +0.000 & \textbf{+0.148} \\
Miniboone & Acc. & 36\% & \textbf{+0.102} & +0.087 \\
Covertype & Acc. & 90\% & +0.008 & \textbf{+0.255} \\
Bioresponse & Acc. & 72\% & +0.060 & \textbf{+0.079} \\
Diabetes & Acc. & 62\% & +0.061 & \textbf{+0.162} \\
Fraud & F1 & 52\% & +0.228 & \textbf{+0.318} \\
Poverty & MAE & 58\% & \textbf{+118} & $-$125 \\
Air & RMSE & 72\% & +2.000 & \textbf{+6.131} \\
Northwind & MAE & 30\% & \textbf{+1.368} & +0.871 \\
\hline
\end{tabular}
\end{table}

\vspace{2mm}
\noindent \textbf{Maximum path length.} 
We vary the maximum join path length $\ell$ from 2 to 7 and assess its impact on effectiveness.
As Table~\ref{tab:sensitivity-max-path-depth} shows, shallow exploration at $\ell=2$ underperforms on eleven of the thirteen datasets, e.g., on \textit{Fraud} (0.443 against 0.573 at $\ell=7$) and \textit{Diabetes} (0.645 against 0.743), because one-hop paths cannot reach the tables that hold the most informative features. 
Allowing a second hop ($\ell=3$) recovers most of this gain, and longer limits change the results little. 
On \textit{Eyemove} and \textit{Fraud}, accuracy peaks at $\ell=3$ and then decreases: longer paths compete with shorter ones for the $\pi=10$ materialized paths and displace informative paths of two and three tables. 
\textit{Northwind}, whose schema chains many-to-one relationships, keeps all its short paths and gains from its longer ones, reaching its best result at $\ell=6$. 
Setting $\ell=7$ as the default thus keeps distant features reachable, while \hippasus still materializes only the top $\pi$ paths.

\vspace{2mm}
\noindent \textbf{Number of Join Paths.} 
We assess the impact of the number $\pi$ of join paths on effectiveness. We vary $\pi$ over 5, 10, and 20, while fixing maximum path length $\ell=7$, $\tau=0.001$, and $\kappa=10$ features selected.
The number of paths trades effectiveness for cost: each additional path is one more join path to materialize and brings more candidate features to score. 
As Table~\ref{tab:sensitivity-num-join-paths} shows, materializing only 5 paths underperforms on eight of the thirteen datasets, as the top five paths often miss tables that hold informative features; \textit{School} and \textit{Poverty} are exceptions, as their first five paths already reach their most informative tables. 
Doubling $\pi$ to 20 improves accuracy on datasets whose features are spread over many tables joined one-to-one with the base table, such as \textit{Jannis}, \textit{Miniboone}, \textit{Covertype}, and \textit{Eyemove}: there, each additional path contributes new columns rather than repeating known ones, and on \textit{Covertype} \hippasus reaches 0.931 at $\pi=20$, above every baseline in Table~\ref{tab:end-to-end-effectiveness}. 
On \textit{Credit}, \textit{Steel}, \textit{Diabetes}, \textit{Air}, and \textit{Northwind}, $\pi=10$ performs best or on par, so additional paths add join cost without improving accuracy. 
We therefore keep $\pi=10$ as the default, which bounds the join cost, while $\pi$ can be raised for schemas with many candidate tables.

\vspace{2mm}
\noindent\textbf{Number of Features Selected.} Varying the number $\kappa$ of selected features (Appendix~\ref{appendix:kappa} at the end of this paper) shows that no single value dominates: retaining more features helps when many features each carry part of the signal, while a few features suffice on datasets such as \textit{Credit} and \textit{Diabetes}. Since $\kappa$ only truncates the final ranking, changing it requires no additional joins. 

\vspace{2mm}
\noindent\textbf{Contribution of multi-hop features.}
To quantify how much of the gain of \hippasus comes from distant tables, we split the augmented features it selects by the number of joins of the path that produced them, retrain on the base features with each group alone, and attribute the gain over Base to the two groups with Shapley values (Table~\ref{tab:multihop}). Features reached through two or more joins form 30--94\% of the selection on all datasets except \textit{School}, whose tables all join directly with the base table, and they contribute more than one-hop features on nine of the thirteen datasets, e.g., on \textit{Covertype}, \textit{Fraud}, and \textit{Diabetes}. Multi-hop exploration thus accounts for a large share of the accuracy gains, which approaches limited to one hop, such as ARDA, cannot obtain.

\section{Conclusion}
\label{sec:conclusions}

We presented \hippasus, a modular framework for effective and efficient feature augmentation in relational databases. 
Our approach addresses the fundamental tradeoff between computational efficiency and feature quality by decoupling the augmentation pipeline into four specialized components: feature description generation, path exploration, join execution, and feature selection. 
We combine lightweight statistical signals with adaptive semantic analysis from Small and Large Language Models to prune unpromising join paths early, employ optimized multi-way join algorithms to efficiently consolidate features from multiple paths, and integrate statistically grounded feature ranking with semantic refinement to identify the most informative features. 
Experiments on publicly available benchmarks demonstrate that \hippasus achieves the strongest overall tradeoff between effectiveness and efficiency. 
For accuracy, \hippasus achieves average improvements of 15.7\% over ARDA, 8.1\% over AutoFeat, and 10.8\% over FeatPilot. 
For runtime performance, \hippasus delivers a median speedup of 64$\times$ over FeatPilot and 3.4$\times$ over ARDA, while maintaining competitive efficiency with AutoFeat. Its adaptive use of language models spends less than half of the LLM input tokens of always invoking the LLM, and most of its gain on nine datasets comes from features two or more joins away, which one-hop methods cannot reach. 
Future work includes extending \hippasus to support data lake settings, where the join graph is not precomputed and may also involve similarity joins.

\section*{Acknowledgment}
\noindent We used an AI assistant (Claude, Anthropic) as a co-pilot for drafting and editing text in all sections and for writing and debugging code. The authors reviewed and verified all of its output and take full responsibility for the content of this paper. The LLMs that \hippasus itself uses are described in Section~\ref{sec:experiments}.

\bibliographystyle{IEEEtran}
\interlinepenalty=10000
\bibliography{references}

\clearpage
\section*{Appendix}
\raggedbottom
\setcounter{subsection}{0}
\renewcommand{\thesubsection}{\Alph{subsection}}
\renewcommand{\theHsubsection}{appendix.\Alph{subsection}}

\subsection{Prompts in \hippasus}
\label{appendix:prompts}

\noindent Here we show the prompts used in \hippasus. 
Figure~\ref{fig:feature-description-generator-prompt} illustrates the prompt used by the Feature Description Generator. 
Furthermore, Figure~\ref{fig:path-explorer-prompt} shows the prompt used in the Path Explorer. 
Finally, Figure~\ref{fig:feature-selector-prompt} displays the prompt used in the Feature Selector. 

\subsection{JEX Correctness and Complexity}
\label{appendix:jex-correctness-and-complexity} 

\begin{theorem}[JEX Correctness]
\label{theorem:JEX-correctness}
Let $p = [T_{\text{base}}, T_1, T_2, \ldots, T_{\ell-1}]$ be a join path and let $D$ be a fixed deterministic deduplication policy applied consistently to all one-to-many or many-to-many joins. 
Then suffix-Yannakakis~$(p, D)$ produces a path materialization result $R_p$ that is equivalent to binary left-join materialization~$(p, D)$ with respect to the rows of~\text{}~$T_{\text{base}}$ and the resulting joined feature values. 
Moreover, $R_p$ satisfies the row-preservation and distribution-preservation invariants by construction. 
\end{theorem}
\begin{proof}[Proof Sketch]
Row preservation follows from the final left outer join onto $T_{\text{base}}$, which preserves every row of the base table. 
Distribution preservation follows immediately, since the target variable $y$ is anchored in $T_{\text{base}}$ and no base-table rows are duplicated or removed. 
Finally, equivalence to binary left-join materialization follows because the base-seeded semi-join reductions remove only suffix tuples whose join key no tuple of the preceding table references, so these tuples cannot appear in any left-join result; the left joins keep every remaining tuple, including those without matches further down the path, and under the same deterministic deduplication policy $D$ both plans keep the same tuple per key and thus yield the same feature values.
\end{proof}

\begin{proposition}[Complexity]
\renewcommand{\theproposition}{}
Let $n = |T_{\text{base}}|$, let $n_i = |T_i|$ for $i \in \{1,\ldots,\ell-1\}$, and let $n_i^* \leq n_i$ denote the size of suffix table $T_i$ after the base-seeded semi-join reduction. With hash-based semi-joins and joins, suffix-Yannakakis materialization runs in expected time
\begin{equation}
O\!\left(n + \sum_{i=1}^{\ell-1} n_i\right),
\end{equation}
and every intermediate result along the suffix has at most $n_1^*$ tuples, whereas the binary strategy materializes $\ell-2$ intermediate results of $n$ tuples each: the reduction scans each suffix table once, the left joins along the suffix never add tuples because each reduced table holds one tuple per key, and the final left outer join touches each base row once.
\end{proposition}

\subsection{Dataset Preprocessing}
\label{appendix:datasets}
\noindent We use the datasets as released with AutoFeat~\cite{icde24-autofeat} and FeatPilot~\cite{icde25-featpilot}, with four corrections that apply identically to every method. \textit{Steel}: the label separates six common fault types from all other faults, and columns V28--V33 are the indicators of exactly these six types (Pastry, Z\_Scratch, K\_Scratch, Stains, Dirtiness, and Bumps), spread over five candidate tables; together they determine the label, so we remove them. \textit{Air}: the target is the daily temperature, and the base-table column \texttt{f0\_} and the column \texttt{NO27} of the daily temperature summaries hold the same near-copy of it ($r=0.997$); we remove both, after which Base has no feature and predicts the training mean. \textit{Poverty}: seven candidate tables contain columns named `Good Feature~1'--`Good Feature~4' that correlate with the target at $|r|=0.97$--$0.98$; we remove them. \textit{Northwind}: the copy used in prior work had lost rows of its \texttt{orders}, \texttt{customers}, \texttt{suppliers}, and \texttt{employees} tables; we restore them from the public Northwind database, leaving the base table unchanged. All methods also drop identifier columns by the same rule. We release the datasets as used.

\subsection{Time Breakdown of \hippasus}
\label{appendix:time-breakdown}
\begin{table}[t]
\centering
\caption{\normalfont Computation time of \hippasus per phase, and its LLM wait, in seconds (default parameters; median of three runs). Total includes loading, consolidation, and splitting.}
\label{tab:time-breakdown}
\small
\setlength{\tabcolsep}{3pt}
\begin{tabular}{lrrrrrr}
\hline
\textbf{Dataset} & \textbf{Rows} & \textbf{PEX} & \textbf{JEX} & \textbf{FS} & \textbf{Total} & \textbf{LLM wait} \\
\hline
School & 1,774 & 1.2 & 0.5 & 0.9 & 2.7 & 17.6 \\
Credit & 1,000 & 0.2 & 0.1 & 0.0 & 0.4 & 7.6 \\
Eyemove & 7,608 & 0.2 & 0.5 & 0.1 & 0.8 & 6.3 \\
Steel & 1,941 & 0.2 & 0.2 & 0.0 & 0.4 & 2.4 \\
Jannis & 57,580 & 1.5 & 2.8 & 1.9 & 6.5 & 0.0 \\
Miniboone & 72,998 & 2.2 & 3.3 & 1.9 & 7.8 & 0.0 \\
Covertype & 423,680 & 9.3 & 35.4 & 14.5 & 61.1 & 3.6 \\
Bioresponse & 3,434 & 0.7 & 0.4 & 0.2 & 1.4 & 0.0 \\
Diabetes & 20,000 & 0.6 & 0.9 & 0.2 & 1.9 & 2.3 \\
Fraud & 20,000 & 0.6 & 1.9 & 0.2 & 2.9 & 6.6 \\
Poverty & 3,136 & 0.2 & 0.2 & 0.4 & 0.8 & 10.8 \\
Air & 11,839 & 0.9 & 1.1 & 1.1 & 3.3 & 8.9 \\
Northwind & 2,155 & 0.2 & 0.1 & 1.1 & 1.4 & 16.7 \\
\hline
\end{tabular}
\end{table}
\noindent Table~\ref{tab:time-breakdown} splits the computation time of \hippasus at the default parameters into its three phases, the Path Explorer (PEX), the Join Executor (JEX), and the Feature Selector (FS), next to its LLM wait; its Total and LLM wait columns are the two parts of each cell of Table~\ref{tab:end-to-end-efficiency}. The computation grows with the number of base-table rows that the joins and the feature selection process: \textit{Covertype}, with 424K rows, spends 35\,s in join execution and 15\,s in feature selection, while every other dataset needs under 8\,s in total. On every dataset with semantic signal except \textit{Covertype}, the LLM wait exceeds the computation; the three datasets without semantic signal make no LLM call.

\subsection{Effect of Feature Description Generation}
\label{appendix:fdg}
\begin{table*}[t]
\centering
\caption{\normalfont Effect of Feature Description Generation (FDG) on the datasets with semantic signal.}
\label{tab:feature-description-ablation}
\small
\setlength{\tabcolsep}{3pt}
\begin{tabular}{l||c|c|c|c|c|c|c|c|c|c}
\hline
\multirow{2}{*}{\textbf{Variant}} & \textbf{School} & \textbf{Credit} & \textbf{Eyemove} & \textbf{Steel} & \textbf{Covertype} & \textbf{Diabetes} & \textbf{Fraud} & \textbf{Poverty} & \textbf{Air} & \textbf{Northwind} \\
& Acc. & Acc. & Acc. & Acc. & Acc. & Acc. & F1 & MAE & RMSE & MAE \\
\hline
\hline
$\hippasus_{\textit{w/o FDG}}$ & \textbf{0.825} & \textbf{0.736} & \textbf{0.617} & \textbf{0.776} & \textbf{0.779} & \underline{0.740} & \textbf{0.582} & \underline{12129} & \textbf{5.743} & \underline{11.829} \\
$\hippasus_{\textit{w/\text{ } FDG}}$ & \underline{0.822} & \textbf{0.736} & \underline{0.577} & \underline{0.775} & \underline{0.767} & \textbf{0.743} & \underline{0.573} & \textbf{12035} & \underline{5.769} & \textbf{11.797} \\
\hline
\end{tabular}
\end{table*}
\begin{table*}[t]
\centering
\caption{\normalfont Effect of the number of selected features.}
\label{tab:sensitivity-features-selected}
\small
\setlength{\tabcolsep}{2pt}
\begin{tabular}{l||c|c|c|c|c|c|c|c|c|c|c|c|c}
\hline
\multirow{2}{*}{\textbf{Features $\kappa$}} & \textbf{School} & \textbf{Credit} & \textbf{Eyemove} & \textbf{Steel} & \textbf{Jannis} & \textbf{Miniboone} & \textbf{Covertype} & \textbf{Bioresponse} & \textbf{Diabetes} & \textbf{Fraud} & \textbf{Poverty} & \textbf{Air} & \textbf{Northwind} \\
& Acc. & Acc. & Acc. & Acc. & Acc. & Acc. & Acc. & Acc. & Acc. & F1 & MAE & RMSE & MAE \\
\hline
\hline
5 & 0.754 & \textbf{0.741} & 0.580 & \underline{0.767} & 0.696 & 0.876 & 0.763 & 0.702 & 0.741 & 0.548 & \textbf{11882} & 6.612 & 11.986 \\
10 & \underline{0.822} & 0.736 & 0.577 & \textbf{0.775} & \underline{0.713} & 0.888 & 0.767 & 0.724 & \textbf{0.743} & \underline{0.573} & 12035 & 5.769 & \textbf{11.797} \\
15 & \textbf{0.825} & 0.738 & \textbf{0.596} & \textbf{0.775} & \textbf{0.715} & \underline{0.903} & \underline{0.778} & \underline{0.727} & \underline{0.742} & \underline{0.573} & \underline{11938} & \underline{5.344} & \underline{11.882} \\
20 & 0.814 & \underline{0.739} & \underline{0.589} & \textbf{0.775} & \underline{0.713} & \textbf{0.910} & \textbf{0.782} & \textbf{0.730} & \underline{0.742} & \textbf{0.585} & 12086 & \textbf{5.145} & 11.884 \\
\hline
\end{tabular}
\end{table*}
\noindent Table~\ref{tab:feature-description-ablation} compares \hippasus with and without FDG on the ten datasets with semantic signal. FDG improves or preserves accuracy on \textit{Credit}, \textit{Diabetes}, \textit{Poverty}, and \textit{Northwind}, whose column names are meaningful: every feature receives a distinct description that restates and clarifies its name (e.g., credit\_amount, BMI). On the other datasets, the differences remain within about one standard deviation across seeds, and they follow the quality of the generated descriptions: on \textit{Air} and \textit{Fraud}, many features share the same description (only 25\% and 48\% of the descriptions are distinct), on \textit{Steel} and \textit{Eyemove}, whose column names are codes or abbreviations, the descriptions must be inferred, and on \textit{Steel} the two configurations are essentially identical. Descriptions taken from dataset documentation are thus a direct way to strengthen this component.

\subsection{Effect of the Number of Selected Features}
\label{appendix:kappa}
\noindent We vary the number $\kappa$ of selected features 
in order to assess its impact on effectiveness only, 
since the Feature Selector's LLM semantically refines the statistical ranking over the consolidated feature pool regardless of how many features are ultimately retained.
The value of $\kappa$ caps how many augmented features the Feature Selector retains, whether it reranks them with the LLM or, on datasets without semantic signal, keeps the top features by mutual information.
Table~\ref{tab:sensitivity-features-selected} presents the effectiveness results, revealing that no single value of $\kappa$ dominates across all datasets. 
Selecting only five features underperforms on most datasets, e.g., on \textit{School} (0.754 against 0.822 at $\kappa=10$), whose signal is spread over several progress-report and survey indicators. 
Retaining more features helps when many augmented features each carry part of the signal, as with the particle measurements of \textit{Miniboone} (0.876 at $\kappa=5$ against 0.910 at $\kappa=20$), the cartographic variables of \textit{Covertype}, and the pollutant measurements of \textit{Air} (RMSE 6.612 against 5.145). 
Conversely, on \textit{Credit}, \textit{Diabetes}, and \textit{Northwind}, a small $\kappa$ suffices, suggesting that a few features carry most of the useful signal. 
Since $\kappa$ only truncates the final ranking, changing it requires no additional joins.

\begin{figure*}[p]
\raggedright
\small
\noindent\rule{\textwidth}{0.4pt}

\noindent\textbf{System:} 
You are a data science expert analyzing database schemas. Given table names, feature names, and optional dataset descriptions, generate concise, semantically-rich descriptions for each feature and determine whether the dataset provides sufficient semantic signal for downstream semantic reasoning. 

\noindent\textbf{Requirements:}  
(1) Generate descriptions for all features provided; 
(2) Each description should be 3-10 words; 
(3) Focus on semantic meaning, not technical data types; 
(4) Transform abbreviations and codes into clear, domain-relevant descriptions; 
(5) In addition, determine whether the dataset provides sufficient semantic signal for downstream semantic reasoning. Set \texttt{semantic\_signal\_available = 1} when feature names and/or generated descriptions convey interpretable domain meaning; otherwise set it to \texttt{0}; 
(6) Return a JSON object of the form: 

\texttt{\{"semantic\_signal\_available": 0/1, "descriptions": \{table\_1.feature1: description\_1, table\_1.feature\_2: description\_2,...\}\}}.

\noindent\textbf{User:} 
Dataset context: \texttt{\textcolor{purple}{\{dataset\_description\}}} (if available). 
Tables and features:
\begin{lstlisting}[basicstyle=\small\ttfamily, escapechar=@, backgroundcolor=]
@\textcolor{purple}{\{table\_1\}}@: [@\textcolor{purple}{\{feature\_1\}}@ (@\textcolor{purple}{\{type\_1\}}@),...], @\textcolor{purple}{\{table\_2\}}@: [@\textcolor{purple}{\{feature\_2\}}@ (@\textcolor{purple}{\{type\_2\}}@),...],...
\end{lstlisting}

\noindent\rule{\textwidth}{0.4pt}
\caption{\normalfont Prompt template for Feature Description Generator.
}
\label{fig:feature-description-generator-prompt}
\vspace{4mm}
\raggedright
\small
\noindent\rule{\textwidth}{0.4pt}

\noindent\textbf{System:} 
You are a data science expert performing semantic table relevance assessment for \texttt{\textcolor{purple}{\{task\_description\}}}. 
Given the base table schema, candidate table schemas, 
feature names and descriptions, and the target variable, score each candidate table by how useful its features are for predicting the target. 

\noindent\textbf{Requirements:}  
(1) Score all candidate tables—no exceptions; 
(2) Score range [0, 100]: 100 = highly relevant, 0 = irrelevant; 
(3) Base your scores on semantic relevance to the prediction task; 
(4) Return a JSON object: 

\texttt{\{table\_1: score\_1, table\_2: score\_2,...\}}.

\noindent\textbf{User:} 
Task: \texttt{\textcolor{purple}{\{task\_description\}}}. 
Target: \texttt{\textcolor{purple}{\{target\_column\}}}.  
Base table: 
\begin{lstlisting}[basicstyle=\small\ttfamily, escapechar=@, backgroundcolor=]
@\textcolor{purple}{\{base\_table\_name\}}@: [@\textcolor{purple}{\{feature\_0\}}@ (@\textcolor{purple}{\{type\_0\}}@, @\textcolor{purple}{\{description\_0\}}@),...]
\end{lstlisting}

Candidate tables: 
\begin{lstlisting}[basicstyle=\small\ttfamily, escapechar=@, backgroundcolor=]
@\textcolor{purple}{\{table\_1\}}@: [@\textcolor{purple}{\{feature\_1\}}@ (@\textcolor{purple}{\{type\_1\}}@, @\textcolor{purple}{\{description\_1\}}@),...], @\textcolor{purple}{\{table\_2\}}@: [@\textcolor{purple}{\{feature\_2\}}@ (@\textcolor{purple}{\{type\_2\}}@, @\textcolor{purple}{\{description\_2\}}@),...],...
\end{lstlisting}

\noindent\rule{\textwidth}{0.4pt}
\caption{\normalfont Prompt template for Path Explorer when the LLM oracle is invoked for semantic table scoring.}
\label{fig:path-explorer-prompt}
\vspace{4mm}
\raggedright
\small
\noindent\rule{\textwidth}{0.4pt}

\noindent\textbf{System:}
You are a data science expert performing feature selection for \texttt{\textcolor{purple}{\{task\_description\}}}. 
The features below are pre-sorted by mutual information (MI) with the target variable, strongest predictor first. 
Your task is to make targeted adjustments to this ranking only when semantic evidence provides a clear reason to override the default MI order. 
Mutual information measures the statistical dependency between a feature and the target; higher MI indicates a stronger measured relationship in the training data. 

\noindent\textbf{Adjustment Rules:}
\begin{itemize}[leftmargin=*,noitemsep]
\item Keep the MI ranking by default.
\item Promote a feature only when its description reveals strong causal or domain relevance that MI may underestimate. 
\item Demote a feature only when there is clear evidence that its high MI is artifactual, such as: 
\begin{itemize}[leftmargin=*,noitemsep]
\item unique identifiers (e.g., names, phone numbers, postal codes, addresses, IDs), which may have high MI simply because they are nearly unique per row; 
\item leakage-prone features collected after the prediction point or directly encoding the target. 
\end{itemize}
\item Do not demote a feature merely because its name is technical or unfamiliar. 
\item When uncertain, preserve the MI ranking. 
\end{itemize}

\noindent\textbf{Requirements:}  
(1) Return all provided features—no exceptions, no omissions; 
(2) Do not add, remove, rename, or modify any feature name; 
(3) Return a JSON array: \texttt{["most\_important",...,"least\_important"]}

\noindent\textbf{User:} 
Task: \texttt{\textcolor{purple}{\{task\_description\}}}. 
Target: \texttt{\textcolor{purple}{\{target\_column\}}}. 
Features with MI-based statistical context (pre-sorted by MI descending): 
\begin{lstlisting}[basicstyle=\small\ttfamily, escapechar=@, backgroundcolor=]
[{"rank": @\textcolor{purple}{\{rank\_1\}}@, "MI": @\textcolor{purple}{\{mutual\_info\_1\}}@, "feature": "@\textcolor{purple}{\{feature\_1\}}@", "description": "@\textcolor{purple}{\{description\_1\}}@"},
 {"rank": @\textcolor{purple}{\{rank\_2\}}@, "MI": @\textcolor{purple}{\{mutual\_info\_2\}}@, "feature": "@\textcolor{purple}{\{feature\_2\}}@", "description": "@\textcolor{purple}{\{description\_2\}}@"},...]
\end{lstlisting}
\noindent\rule{\textwidth}{0.4pt}
\caption{\normalfont Prompt template for Feature Selector with MI-anchored semantic reranking.}
\label{fig:feature-selector-prompt}
\end{figure*}
\clearpage

\end{document}